\documentclass[10pt,twoside]{article}

\usepackage{fancybox,amsfonts} 
\usepackage{graphicx}

\usepackage{amssymb}
\usepackage{latexsym}

\makeatletter

\@addtoreset{equation}{section} \makeatother
\newtheorem{theorem}{Theorem}[section]
\newtheorem{remark}[theorem]{Remark}
\newtheorem{lemma}[theorem]{Lemma}
\newtheorem{proposition}[theorem]{Proposition}
\newtheorem{corollary}[theorem]{Corollary}
\newtheorem{definition}[theorem]{Definition}

\newcommand{\cvd}{\ \rule{0.5em}{0.5em}}
\newcommand{\be}{\begin{equation}}
\newcommand{\ee}{\end{equation}}

\newcommand{\R}{{\mathbb R}}
\newcommand{\LL}{{\mathbb L}}
\newcommand{\SSS}{{\mathbb S}}

\newcommand{\M}{{\cal M}}

\newcommand{\C}{{\cal C}}
\newcommand{\J}{{\cal J}}

\newcommand{\noi}{\noindent}
\newcommand{\ben}{\begin{enumerate}}
\newcommand{\een}{\end{enumerate}}
\newcommand{\bit}{\begin{itemize}}
\newcommand{\eit}{\end{itemize}}
\newcommand{\edoc}{\end{document}}

\newcommand{\np}{\newpage}

\title{
\vspace{0.5in} {\bf  The causal boundary of wave-type spacetimes}}

\author {\bf J.L. Flores$^{1}$, M.
S\'anchez$^{2}$\thanks{Both authors partially supported by Spanish
MEC-FEDER Grant MTM2007-60731 and Regional J. Andaluc\'{\i}a
Grant P06-FQM-01951. JLF  also  supported by MEC Grant RyC-2004-382.} \\
$^1$ {\it\small Departamento de \'Algebra, Geometr\'{\i}a y Topolog\'{\i}a}\\
{\it\small Facultad de Ciencias, Universidad de M\'alaga} \\ {\it\small Campus Teatinos, 29071 M\'alaga, Spain} \vspace{1mm}\\
$^2$ {\it\small Departamento de Geometr\'{\i}a y Topolog\'{\i}a}\\
{\it\small Facultad de Ciencias, Universidad de Granada}\\
{\it\small Avenida Fuentenueva s/n, 18071 Granada, Spain} }

\begin{document}

\setlength{\shadowsize}{2pt}

\newcommand{\scaja}[3]{
\begin{minipage}[c][#1]{14.5em}
\begin{center}\shadowbox{
\begin{minipage}[c][#1]{#2}
#3
\end{minipage}
}\end{center}
\end{minipage}
}

\newcommand{\caja}[3]{
\begin{minipage}[c][#1][c]{10em}\begin{center}
\begin{minipage}[c][#1][c]{#2}
#3
\end{minipage}\end{center}
\end{minipage}
} 

\parindent=5mm
\date{}
\maketitle

\begin{quote}

\noindent {\small \bf Abstract.} {\small A complete and systematic
approach to compute the causal boundary of wave-type spacetimes is
carried out. The case of a 1-dimensional boundary is specially
analyzed and its critical appearance in pp-wave type spacetimes is
emphasized. In particular, the corresponding results  obtained in
the framework of the AdS/CFT correspondence for holography on the
boundary, are reinterpreted and very widely generalized.

Technically, a recent new definition of causal boundary is used
and stressed. Moreover, a set of mathematical tools is introduced
(analytical functional approach, Sturm-Liouville theory,
Fermat-type arrival time, Busemann-type functions).}\\

\end{quote}
\begin{quote}
{\small\sl Keywords:} {\small causal boundary, causal structure,
conformal boundary, pp-waves, plane fronted waves, Mp-waves,
spacetime functional approach, Fermat's principle, plane
wave string backgrounds, Penrose limit, AdS/CFT.}\\

\end{quote}
\begin{quote}

{\small\sl 2000 MSC:} {\small 53C50, 83E30, 83C35, 81T30.}
\end{quote}

\newpage
{\small \tableofcontents } \np


\section{Introduction}

There are several motivations for the recent interest on the
boundary of wave type spacetimes. Firstly, there are important
reasons for string theory, because of the AdS/CFT correspondence
of plane waves and the holographic role of its boundary. But there
are also reasons from the viewpoint of  General Relativity, apart
from the obvious interest in the properties of a classical
spacetime. In fact, the old problem on the consistency of causal
boundaries and its relation with conformal boundaries is put
forward by pp-waves and stimulates its full solution. Very
roughly, the  main results can be summarized as follows (see also
references therein):

\bit \item Plane waves yield exact backgrounds for string theory
as all their scalar curvature invariants vanish. Thus, they
correspond to exact conformal theories, and in some cases can be
explicitly quantized \cite{AK, HS, Me}. \item Taking into account
the well-known result that any spacetime has a plane wave as a
limit along any lightlike geodesic (Penrose, \cite{Pe}),
Berenstein, Maldacena and Nastase \cite{BMN} related string theory
on maximally supersymmetric 10 dimensional plane waves to 4
dimensional field theory.  \item More precisely, Penrose limit on
a lightlike geodesic on AdS$_5\times S^5$, which rotates on the
$S^5$ was considered. Blau, Figueroa-O'Farrill, Hull and
Papadopoulos \cite{BFHP} constructed the limit plane wave and
identified its dual in the field theory. Berenstein and Nastase
\cite{BN} studied the asymptotic {\em conformal} boundary of this
plane wave, finding that it is 1-dimensional. This fact not only
was not regarded as  pathological, but it suggested that such a
plane wave possesses a holographic dual description in terms of
quantum mechanics on its boundary --a similar picture to
 CFT dual to an asymptotically AdS space. \item Marolf and Ross
\cite{MR} studied the {\em causal} boundary of that plane wave.
There are interesting reasons to use this more sophisticated
boundary. On one hand, it is intrinsic to the spacetime and
systematically determined. On the other, this approach is
applicable to any plane wave or spacetime, not only to conformally
flat ones. Essentially, these authors reobtained the 1-dimensional
character for the causal boundary of Berenstein and Nastase's,
and, {\em surprisingly}, obtained other relevant cases of plane
waves with this same behavior (as it was independent on the number
of positive eigenvalues for the quadratic form $F$, assuming the
existence of at least one). What is more, their results suggested
a redefinition of classical causal boundary \cite{MR2}, as this
old concept was known to have some undesirable properties. \item
In \cite{FS} the authors studied systematically the causal
structure of wave-type spacetimes (the general family
(\ref{m-general}) below). We showed that this structure depends
dramatically on the value of the characteristic coefficient $F$ of
the metric. In particular, when $F$ is ``at most quadratic'' (as
in classical plane waves) the spacetime becomes strongly causal,
but when it is ``superquadratic'' the wave is non-distinguishing
and the causal boundary makes no sense. Hubeny, Rangamani and Ross
\cite{HRR} pointed out that this is the case of the pp-wave which
gives rise to the ${\cal N}=2$ sine-Gordon string world-sheet;
moreover, they also studied other properties on causality (as the
existence of time functions) and boundaries for some specific
pp-wave backgrounds \cite{HRR, HR, HRRa, HRRb}. \eit

\noi There are also two technical questions which are worth of
pointing out here. First, the systematic study of the causal
boundary in \cite{F}, starting at the cited original idea
\cite{MR2}, which seems to yield a definitive answer to the
problem of the identifications between future and past ideal
points, as well as an appropriate topology on the boundary.
Second, the solution of the so-called ``folk problems of
smoothability'' which yield consistency to the full causal ladder
of causality, including the equivalence between stable causality
and the existence of a time function \cite{BeSa, MS, BeSa2}.

\smallskip

\noindent The aim of the present article is to study
systematically the causal boundary of wave-type spacetimes. Recall
that, essentially, Marolf and Ross \cite{MR, MR2} studied  locally
symmetric plane waves ($F(x,u)\equiv F(x)$, $F$ quadratic form),
and Hubeny and Rangamani \cite{HR} studied particular cases of
plane waves, as well as some pp-waves, extracting some heuristic
conclusions. But more precise and general results about the
structure of the boundaries are missing there.

Summing up, our motivation is threefold: first to conclude the
study in \cite{MR, HR}, originated by applications on strings,
second to conclude the study of causality of pp-wave type
spacetimes initiated in \cite{FS, CFS}, and third to check and
support the new concept of causal boundary in \cite{MR2, F}.
Our approach can be summarized as follows.

In Section \ref{s2} we introduce the general class of wave-type
spacetimes, namely Mp-waves ${\cal M}=M\times \R^2$, to be
considered. Other properties of these spacetimes (geodesics,
completeness, causal hierarchy) were studied in \cite{CFS, FS};
some changes of notation are made here.

In Section \ref{s3} the framework of causal boundaries is
introduced. First, the original Geroch, Kronheimer and Penrose
(GKP) boundary of TIP's and TIF's \cite{GKP} is recalled
$\S$\ref{s3.1}. The  recent progress on this boundary \cite{MR2,
F} applicable here is summarized in $\S$\ref{s3.2}. This includes
the characterization of ideal points as certain pairs $(P,F)$ of
TIP's and TIF's (which involves their common futures and pasts
$\uparrow P, \downarrow F$),  the induced causal relation and the
topology of the boundary. Finally, a simple, but general,
technical property of TIP's and TIF's is proved in $\S$\ref{s3.3}.
Essentially, this property means that TIP's and TIF's can be
regarded as pasts or futures of certain (non necessarily geodesic)
inextendible lightlike curves (Prop. \ref{plightlike}); its
version for Mp-waves (Cor. \ref{cLIGHT}) will simplify the
functional approach to be used later.

In Section \ref{s4} we introduce an arrival time function with
analogies to classical Fermat's one \cite{Perlick}. This function
allows to introduce a functional ${\cal J}_{u_0}^{\Delta u}$ in
the space of curves on the spatial $M$ part (essentially, in the
set of curves $x(u)$ which connect each two prescribed points
$x_0, x_1 \in M$ parametrized by the ``$u$-quasitime'' $u\in
[u_0,u_0+\Delta u]$, where $(x,u,v)\in M\times \R^2$). The infimum
of ${\cal J}_{u_0}^{\Delta u}$ characterizes which points can be
causally joined with each $(x_0,u_0,v_0) \in \M$. This approach,
on one hand, allows to introduce techniques and results from
functional analysis (some required ones will be developed in the
Appendix). On the other, clarifies the causal structure of
Mp-waves; for example, the inexistence of horizons (claimed in
\cite{HR2} and strongly supported in \cite{FShonnef}) becomes now
apparent (Remark \ref{re-4.2}).

In Section \ref{s5} we introduce two technical conditions {\bf
(H1)}, {\bf(H2)} on the Mp-wave in terms of functional ${\cal J}$
(Defn. \ref{3.3}), and relate them to the qualitative behavior of
the characteristic metric coefficient $F$. Very roughly, the idea
is as follows. Each $M$-curve $x$ determines univocally a
lightlike curve type $(x(u),u,v(u)), u\in [u_0,u_0+\Delta u]$.
Assume that the lightcones become opened fast along the lightlike
curves generated in one $M$-direction (or even just along a
sequence $\{x_m\}_m$ of $M$-loops). Due to the structure of the
Mp-wave, if this happens for arbitrarily small values of $\Delta
u$ (as formally expresses {\bf (H2)}) then the future of all the
points $(x,u,v)$ with the same $u=u_0$ collapses. So, the Mp-wave
will be non-distinguishing, and no causal boundary can be defined.
Now, assume that the Mp-wave is causally well-behaved and, so,
this property does not hold for arbitrarily small $\Delta u$. If
the property still holds for values of $\Delta u$ greater than
some constant $\Delta_0 >0$ (as expresses {\bf (H1)}), then the
collapse will happen at the level of the TIP's, i.e.: lightlike
curves with unbounded coordinate $u$ will generate the same ideal
points $i^{+}$, $i^{-}$.

As conditions  {\bf (H1)}, {\bf(H2)} are formulated directly on
the functional, they become very technical. Nevertheless, we also
define the typical behaviors of $F$ at infinity: super, at most,
and sub quadratic (these are general bounds on the growth of
$F(\cdot , u)$, depending arbitrarily on $u$) as well as
$\lambda$-asymptotically quadratic (such a bound is also
restrictive on $u$). We showed
in \cite{FS} how some of these behaviors 
determine the position in the causal ladder of the Mp-wave. Now,
we show (Lemmas \ref{l11}, \ref{l22}) how some of them
(superquadratic, $\lambda$-asymptotically quadratic with
$\lambda>1/2$) yield naturally conditions {\bf (H2)}, {\bf(H1)},
which will determine its boundary. The results are very accurate,
as shown by the bound $\lambda>1/2$, which comes from
Sturm-Liouville theory  (see Remark \ref{rsturm} and
$\S$\ref{s9.1}). Nevertheless, we emphasize that the technical
behavior {\bf (H1)},
{\bf(H2)} is required only for some $M$-direction. 
 Thus, one can easily yield
results more general than stated. In fact, in Lemma \ref{l22}(ii)
condition {\bf (H1)} is proved for (non-necessarily locally
symmetric)
 plane waves such that one of the eigenvalues of $F$ is positive; this lies in the
 core of the  surprising
result by Marolf and Ross \cite{MR} cited above.

In Section \ref{s6} we prove how {\bf(H2)} forbids the Mp-wave to
be distinguishing ($\S$\ref{s6.1}). This may be somewhat
unexpected, and  some examples in \cite{HR} are revisited
($\S$\ref{s6.2}).

In Section \ref{s7} the explicit construction of the ideal points
for any strongly causal Mp-wave is carried out. This is done in
full generality in $\S$\ref{s7.1}, where the main result (Theorem
\ref{t-pastsets}) is expressed in terms of two ``Busemann type
functions'' $b^\pm$ previously introduced (Props. \ref{l-I-},
\ref{l-I-f}). Notice that Busemann functions appear naturally when
TIP's  or TIF's  are computed in simple (standard static)
spacetimes, \cite{Harris}. Now, the more elaborated function $b^-$
plays a similar role to such a Busemann function, and the new
function $b^+$ is introduced to deal with the sets $\uparrow P,
\downarrow F$ required for the total causal boundary \cite{MR2,
F}. Moreover ($\S$\ref{s7.2}), when $|F|$ is at most quadratic
(and, thus, $\M$ is necessarily strongly causal) and $M$ complete,
a special simplification of the terminal sets $P, F, \uparrow P,
\downarrow F$ occurs. (We emphasize the necessity of the at most
quadratic behavior for $|F|$, which was dropped in previous
literature, Remark \ref{rpalo}.) In fact, a natural lightlike
ideal line in each boundary $\hat{\partial} \M$, $\check{\partial}
\M$ (parametrized by $u_\infty, |u_\infty|<\infty$ in Th. \ref{3},
Remark \ref{r1}) appears.
Nevertheless, the boundary $\partial \M$ may be higher
dimensional, because the lightlike curves with unbounded
coordinate $u$ ($|u_\infty|=\infty$) may still generate infinitely
many ideal points.

However, in Section \ref{s8} we show that, when additionally {\bf
(H1)} holds, then all (future) lightlike curves with
$u\nearrow\infty$ generate the same ideal point $i^+$, so  a
1-dimensional boundary is expected ($\S$\ref{s8.1}). In
particular, {\em when $F$ is $\lambda$-asymptotically quadratic
with $\lambda>1/2$ the boundary is two copies of a 1-dimensional
lightlike line, with some eventual identifications}
($\S$\ref{s8.2}). Moreover the special case of plane waves is
compared carefully with previous results and techniques (Remark
\ref{rcagna} and below).

In Section \ref{s9} we consider subquadratic $F$'s and emphasize
the critical character of the 1-dimensional boundary. Recall that,
essentially, such a boundary corresponds to a $(\lambda
> 1/2)$-asymptotically quadratic behaviour of $F$, and the boundary
makes no sense under a faster (superquadratic) growth. In
$\S$\ref{s9.1} we construct an explicit example with higher
dimensional boundary in the limit case $\lambda=1/2$. So, the
1-dimensional boundary can no longer be expected.

Higher dimensionality is expected specially in the (globally
hyperbolic) subquadratic case $\S$\ref{s9.2}. Notice  that the
case $M=\R^n, F\equiv 0$ corresponds to Lorentz-Minkowski
$\LL^{n+2}$ (for arbitrary $M$, corresponds to a standard static
spacetime). If $|F(\cdot, u)|$ is upper bounded for each $u$, then
the spacetime becomes ``isocausal'' (in the sense of
Garc\'{\i}a-Parrado and Senovilla \cite{GP-Se1}) to $\LL^{n+2}$
and, thus, the causal boundary is expected to be
$(n+1)$-dimensional.

Finally, in $\S$\ref{s9.3} we discuss and extend  Marolf and Ross'
result \cite[Sect. 3.1]{MR} for plane waves with negative
eigenvalues. Concretely, we reobtain that  the Mp-wave is
conformal to a region of $\LL^{n+2}$ bounded by two lightlike
hyperplanes, even when $F$ depends on $u$. Nevertheless, a
 discussion shows that the causal and conformal boundaries differ
in this case: the former has two connected pieces (a future
boundary and a past one); the latter, which is necessarily
compact, is connected and includes implicitly properties at
spacelike infinity (compare with \cite{MR3}).

We finish emphasizing some conclusions in Section \ref{s10},
including a table of results, and providing some technical bounds
on some functionals in the Appendix. Along the paper, four figures
have been also included as a guide for the reader.

\section{Wave-type spacetimes}\label{s2}

\noindent The authors, in collaboration with A.M. Candela,
introduced and studied systematically \cite{CFS, FS, FShonnef} the
following class of spacetimes, which widely generalize classical
pp-waves (and, thus, plane waves):
\begin{equation}\label{m-general}
\begin{array}{c}
(\M,\langle\cdot,\cdot\rangle_{L})\qquad \M= M \times \R^{2}
\\ \langle\cdot,\cdot\rangle_{L}
=\langle\cdot,\cdot\rangle -F(x,u)\,du^{2} - 2\,du\,dv.
\end{array}
\end{equation}
Here $(M,\langle\cdot,\cdot\rangle)$ is any smooth Riemannian
($C^{\infty}$, positive-definite, connected) $n$-manifold, the
variables $(u,v)$ are the natural coordinates of $\R^{2}$ and
$F:M\times\R\rightarrow\R$ is any smooth scalar field. $M$ will
not be assumed to be complete a priori, and will be said {\em
unbounded} if it is non-compact with points at arbitrary long
distances (i.e., it has infinite diameter).

These spacetimes were named just PFW (``plane fronted waves'') in
some previous references but, according to the more careful
notation in the survey \cite{GP-Se}, they will be considered as (a
type of) {\em Mp-waves}. We also introduce some changes of
conventions and notations in order to make a better comparison
with references such as \cite{MR, HR, Sa-Bari}. In particular,
function $F$ here replaces $-H$ in previous references. We will
choose once for ever a point $\bar x \in M$. Then, if $d$ is the
natural distance associated to the Riemannian metric $\langle
\cdot , \cdot \rangle$, we put \be \label{dist} |x|= d(x, \bar x)
\quad \forall x\in M. \ee

Elementary properties of these spacetimes are the following.
Vector field $\partial_{v}$ is parallel and lightlike, and the
time-orientation will be chosen to make it future-directed. Thus,
for any future-directed causal curve $\gamma(s)=(x(s), u(s),
v(s))$, $s\in I$ ($I$ interval)
\begin{equation}\label{gg}
\dot{u}(s)=-\langle\dot{\gamma}(s),\partial_{v}\rangle_{L}\geq 0,
\end{equation}
being the inequality strict if $\gamma(s)$ is timelike (and
analogously for a past-directed curve). Using this inequality and
the fact that $\nabla u=-\partial_{v}$, it follows that any such
Mp-wave is causal. The slices $u\equiv\,$constant are degenerate,
with radical Span$\,\partial_{v}$. Then, all the hypersurfaces
(non-degenerate $n$-submanifolds of $\M$) of one such a slice
which are transverse to $\partial_{v}$, become isometric to open
subsets of $M$. The fronts of the wave (\ref{m-general}) will be
defined as the (whole) slices at constant $u,v$.

\section{The Causal Boundary of spacetimes}\label{s3}

We refer to well-known references such as \cite{O, BEE, HE,  W}
and specially the recent review \cite{MS} for notation and
background on causality. For the specific  approach on causal
boundaries, we refer to \cite{F} and references therein.

\subsection{Classical approach}\label{s3.1}

Let $\M\equiv (\M,g)$ be a spacetime, endowed with a
time-orientation (implicitly assumed) and, thus, the causal $\leq$
(strict causal $<$) and chronological $\ll$ relations. As usual,
causal elements in any open subset $U\subseteq \M$, regarded as a
spacetime in its own right, will be denoted such as $<_U,$
$J^+(p,U)$, etc. A continuous curve $\gamma:[0,b)\rightarrow \M$
is called {\em future-directed causal} if, for each $s\in [0,b)$,
there exists a convex neighborhood (i.e. a (starshaped) normal
neighborhood of all its points) $U$ of $\gamma(s)$ such that,
whenever $s' \in (s,b)$ (resp. $s'\in [0,s)$) satisfies that
$\gamma([s,s'])$ (resp. $\gamma([s',s])$) is included in $U$, then
$\gamma(s)<_U \gamma(s')$ (resp. $\gamma(s')<_U \gamma(s)$). It is
well-known that, up to a reparametrization, such curves are
locally Lipschitzian as well as other properties
\cite[Appendix]{CFS2}, \cite[Sect. 3.5]{MS}. This definition (and
related properties) are naturally extended not only to the past
case, but also to other domains for $\gamma$ different to $[0,b)$;
definitions are also extended to {\em timelike} curves, with no
further mention. A (future or past-directed) causal curve
$\gamma:[0,b)\rightarrow \M$ is {\em piecewise smooth} if there
exists a sequence $\{s_i\} \nearrow b$, $s_0=0$ such that $\gamma$
is smooth on each interval $[s_i,s_{i+i}]$ for all $i$. Notice
that, at any (possibly non-smooth) {\em break} $\gamma(s_i), i>0$,
there are two limit derivatives $\dot{\gamma}(s_i^-)$,
$\dot{\gamma}(s_i^+)$, which are causal vectors in the same cone.
A piecewise smooth geodesic will be called a {\em broken}
geodesic.

Roughly, the main purpose of the causal completion of a spacetime
is to make inextendible timelike curves to end at some
point\footnote{In this sense, the name of {\it chronological
completion} would be more appropriate (as in \cite{F}).
Nevertheless, here we will maintain the term {\it causal
completion} to emphasize that some causal elements have been
introduced, and in close correspondence with previous literature
such as \cite{MR,HR}.}. So, `ideal points' are added to the
spacetime, in such a way that any timelike curve has some endpoint
in the new extended space (at the original manifold or at an ideal
point). To this aim, there will not be any difference if the
(timelike) curves are required to be smooth, piecewise smooth or
continuous. So, in what follows, all the curves will be piecewise
smooth, except when otherwise is said explicitly. The natural
level in the causal hierarchy of spacetimes required for the
completion of $(\M,g)$ is {\em strong causality}. In fact, to be
(pointwise future or past) distinguishing will be a minimum
property in order to recover the points of $\M$ from the general
construction, but strong causality will be necessary to recover
the topology too, as well as for other technical properties.

In order to describe the completion procedure some terminology is
required first.  A subset $P\subseteq \M$ is called a {\em past
set} if it coincides with its chronological past $I^-[P]$,  that
is, $P=I^{-}[P]:=\{p\in \M: p\ll q\;\hbox{for some}\; q\in P\}$.
Given a subset $S\subseteq \M$, we define the {\em common past} of
$S$ as $\downarrow S:=I^{-}[\{p\in \M:\;\; p\ll q\;\;\forall q\in
S\}]$. Notice that $I^{-}[P]$ is always open, and we have chosen
the definition of $\downarrow S$ in order to make it open too. A
non-empty past set that cannot be written as the union of two
proper subsets, both of which are also past sets, is called {\em
indecomposable past} set, IP. An IP which does coincide with the
past of some point in $\M$ is called {\em proper indecomposable
past set}, PIP and, otherwise, {\em terminal indecomposable past
set}, TIP. Of course, by replacing the word `past' by `future' we
obtain the corresponding notions for {\em future set}, {\em common
future}, IF, PIF and TIF.

To construct the future causal completion, firstly identify every
event $p\in \M$ with its PIP, $I^{-}(p)$. Then, define the {\em
future causal boundary} $\hat{\partial}\M$ of $\M$ as the set of
all TIPs in $\M$. Therefore, {\em the future causal completion}
$\hat{\M}$ becomes the set of all IPs:
\[
\M\equiv \hbox{PIPs},\qquad \hat{\partial}\M\equiv
\hbox{TIPs},\qquad \hat{\M}\equiv \hbox{IPs}.
\]
Analogously, every event $p\in \M$ can be identified with its PIF,
$I^{+}(p)$, then the {\em past causal boundary}
$\check{\partial}\M$ of $\M$ is the set of all TIFs in $\M$ and
thus, {\em the past causal completion} $\check{\M}$ is the set of
all IFs:
\[
\M\equiv \hbox{PIFs},\qquad \check{\partial}\M\equiv
\hbox{TIFs},\qquad\check{\M}\equiv \hbox{IFs}.
\]

In order to define the (total) causal completion, the space
$\hat{\M}\cup\check{\M}$ appears obviously. However, it becomes
evident that, in order to obtain a reasonably consistent
definition: (a) PIP's and PIF's must be identified in an obvious
way ($I^{-}(p)\sim I^{+}(p)$ on $\hat{\M}\cup\check{\M}$ for all
$p\in \M$), and (b) the resulting space $\M^{\sharp}$ does not
provide a satisfactory description of the boundary of $\M$,
because this procedure often attaches two ideal points where we
would expect only one (consider the boundary for  the interior of
a $(n-1)$-rectangle in Lorentz-Minkowski $\LL^{n}$: each point at
any timelike side determines naturally both, a TIP and a TIF).
There have been many attempts to define additional identifications
between elements of $\hat{\partial}\M\cup\check{\partial}\M$ in
order to overcome this problem \cite{GKP, BS, R, S1, S2}, but
without totally satisfactory results up to now.

\begin{figure}[h]
\begin{center}
\includegraphics[width=13cm]{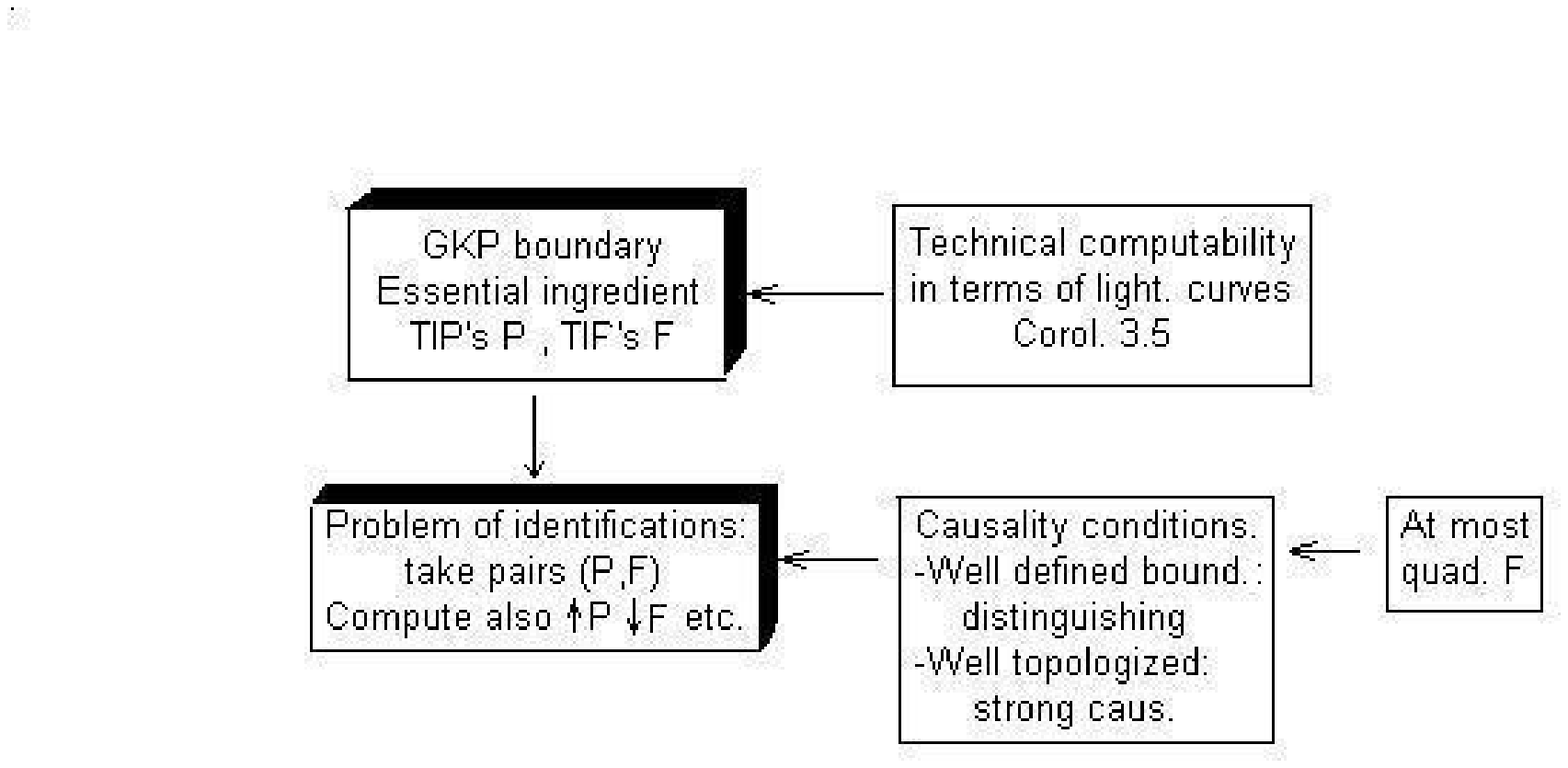}
\end{center}
\caption{Overall causality framework}
\end{figure}

\subsection{A recent new approach}\label{s3.2}

An alternative procedure to making identifications consists of
forming pairs composed by past and future indecomposable sets of
$\M$. This approach, firstly introduced  by Marolf and Ross
\cite{MR2}, and widely developed in \cite{F}, has exhibited
satisfactory results for the spacetimes analyzed up to date, and
seems specially well-adapted to those ones analyzed in
\cite{MR,HR}; so, we will adopt this approach in this paper. Even
though, as emphasized in \cite{F}, there are different choices for
the meaning of the total causal boundary once the pairs have been
defined, they coincide in most cases and, in particular, in the
relevant cases considered here.

Let $P$ (resp. $F$) be an IP (resp. IF). We say that $P$ is {\em
S-related} (Szabados related) to $F$, namely $P\sim_{S}F$, if $P$
is maximal as IP into $\downarrow F$ and $F$ is maximal as IF into
$\uparrow P$. A TIP $P$ can be S-related with more than one TIF
$F_1, F_2$ (take $P=\{(x,t): |x|<-t\}$ in $\M=\LL^2\backslash
\{(0,t): t\geq 0\}$) or viceversa. Nevertheless, this will not
happen in our study (Remark \ref{r1}). Therefore, according to
\cite{F, MR2}, the {\em (total) causal completion $\overline{\M}$}
is defined in this case as: the set of pairs $(P,F)$ where $P$
(resp. $F$) is either a IP (resp. IF) or the empty set and one of
the following possibilities happens: (a) $P\sim_{S}F$, (b) $F\neq
P=\emptyset$ and there is no $P'$ such that $P'\sim_{S}F$, or (c)
$P\neq F=\emptyset$ and there is no $F'$ such that $P\sim_{S}F'$.
The {\em (total) causal boundary} is the subset $\partial \M
\subset \overline{\M}$ containing the pairs $(P,F)$ such that $P$
is not a PIP (and, thus, $F$ is not a PIF \cite[Prop. 5.1]{S1}).

With this definition at hand, it is easy to extend the
chronological relation $\ll$ to the completion $\overline{\M}$:
$(P,F)$ is {\em chronologically related} to $(P',F')$, namely
$(P,F)\ll (P',F')$, if $F\cap P'\neq\emptyset$. The properties and
absence of problems for this choice are well established \cite{F};
nevertheless, it is not so easy to give a definitive extension of
the causal relation. As the boundary of some waves is sometimes
claimed to be null in a rather intuitive way, we will adopt here
simple definitions which will formalize this, and postpone to
future work other subtleties in more general cases. We say that
$(P,F)$ is {\em  causally related} to $(P',F')$, namely $(P,F)\leq
(P',F')$, if  $F'\subseteq F$ and $P\subseteq P'$, at least one of
them not trivially  (i.e., without involving the empty set, $P\neq
\emptyset$ or $F'\neq \emptyset$). This is a canonical choice to
define a causal relation from a chronological one (taken also in
\cite{MR2}; see \cite[Defn. 2.22, Th. 3.69]{MS} for a discussion).
If we only impose that one of these two inclusions hold (not
trivially), we will say $(P,F)$ is {\em weakly causally related to
} $(P',F')$, written $(P,F)\leq_w (P',F')$. It is easy to check
that the latter definition does not imply the former one (consider
in $\LL^2\backslash\{(x,t)\in \R^2: x\leq 0\}$ the ideal point
associated to $(0,0)$ and the pair associated to the point
$(-1,1)$). Finally, $(P,F)$ and $(P',F')$ are {\em (weakly)
horismotically related}
if they are (weakly) causally, but not chronologically, related.

The topology of the spacetime can be also extended to the
completion. We will adopt the {\em chronological topology}
introduced in \cite{F}. This topology is defined in terms of the
following {\em limit operator} $L$: given a sequence
$\sigma=\{(P_{n},F_{n})\}\subset\overline{\M}$ and $(P,F)\in
\overline{\M}$, we say that $(P,F)\in L(\sigma)$ if\footnote{By LI
and LS we mean the usual inferior and superior limits of sets:
i.e. LI$(A_{n})\equiv
\liminf(A_{n}):=\cup_{n=1}^{\infty}\cap_{k=n}^{\infty}A_{k}$ and
LS$(A_{n})\equiv
\limsup(A_{n}):=\cap_{n=1}^{\infty}\cup_{k=n}^{\infty}A_{k}$. This
definition is naturally extended when the range of $n$ is not the
set of natural numbers, but a totally ordered set such as the
interval $[0,b)$ (this permits to say when a curve $\gamma$ tends
to a boundary point $(P,F)$ without using sequences).}
\[
\begin{array}{c}
 P\in\hat{L}(P_{n}):=\{P'\in\hat{\M}: P'\subseteq {\rm
LI}(P_{n})\;\;\hbox{and}\;\; P'\;\;\hbox{is maximal IP into}\;\; {\rm LS}(P_{n})\} \\
F\in\check{L}(F_{n}):=\{F'\in\check{\M}: F'\subseteq {\rm
LI}(F_{n})\;\;\hbox{and}\;\; F'\;\;\hbox{is maximal IF into}\;\;
{\rm LS}(F_{n})\}
\end{array}
\]
(recall that either $P$ or $F$ can be empty, but not both). Then,
the {\em closed sets} for the chronological topology are those
subsets $C\subseteq \overline{\M}$ such that $L(\sigma)\subseteq
C$ for any sequence $\sigma$ in $C$.


\subsection{TIP's as past of lightlike curves}\label{s3.3}

In order to study the pairs in $\partial\M$, it is well-known that
any TIP, $P$, of a strongly causal spacetime can be regarded as
$I^-[\rho]$ for some inextendible future-directed timelike curve
$\rho$ (see, for example, \cite[Prop. 6.14]{BEE}) and analogously
for TIF's. Moreover, in this case, it is $\uparrow \rho = \uparrow
I^-[\rho].$ Our aim is to show that lightlike broken geodesics are
also enough.

\begin{remark} \label{r-contraej} {\em As a previous technicality, recall that when $\gamma$ is
lightlike, then  $\uparrow\gamma=\uparrow I^-[\gamma]$ does not
 necessarily hold: take $\gamma(s)= (s,s), s<0$ in $\LL^2\backslash \{(0,t):
t\geq 0\}$). }\end{remark} Nevertheless, this property is ensured
when the easily checkable condition $\gamma \subset I^-[\gamma]$
holds.

\begin{proposition}\label{ded} Let $\gamma: [0,b) \rightarrow \M$ be
a future-directed  (right) inextendible lightlike curve in the
strongly causal spacetime $\M$. If $\gamma \subset I^-[\gamma]$
then $P=I^-[\gamma]$ is a TIP and $\uparrow\gamma=\uparrow P$.
\end{proposition}

{\it Proof.} Take a sequence $\{s_i\} \nearrow b$. The assumption
on $\gamma$ implies the existence of a subsequence $\{s_{i_k}\}_k$
such that $\gamma(s_{i_k}) \ll \gamma(s_{i_{k+1}})$ for all $k$.
Thus, joining each pair of points by means of a future-directed
timelike curve, a piecewise smooth inextendible timelike curve
$\rho$ is obtained. Clearly, $I^-[\rho]= I^-[\gamma]$ (and thus, a
TIP), $\uparrow \gamma = \uparrow \rho (=\uparrow I^-[\rho])$, and
the result follows. \cvd

\begin{proposition} \label{plightlike}
Let $\rho: [0,b)\rightarrow \M$ be a future-directed causal curve.
Then, for any sequence $\{s_i\} \nearrow b$, $s_0 \geq 0$ there
exists a broken future-directed lightlike geodesic (with no
conjugate points in each unbroken piece) $\gamma: [0,b)\rightarrow
\M$ such that $\gamma(s_i)=\rho(s_i)$ for all $i$ and, thus:
\[
I^-[\rho]= I^-[\gamma], \quad \quad \uparrow \rho = \uparrow
\gamma.
\]
Even more, if (a) dim$\,\M\geq 3$, (b) $X$ is any lightlike
geodesic vector field and (c) the restriction of $\rho$ to any
open interval is not an integral curve of X (up to
reparametrization), then $\gamma$ can be chosen such that
$\dot{\gamma}(s)$ is linearly independent of $X_{\gamma (s)}$ for
all $s\in [0,b)$. In particular, this holds if $\rho$ is timelike;
moreover, in this case $\gamma\subset I^-[\gamma]$.
\end{proposition}
For the proof, notice first:

\begin{lemma} \label{llightlike}
For each $s\in [0,b)$ (resp. $s\in (0,b)$) there exists some
$\epsilon >0$ such that, if $s'\in (s,s+\epsilon)$ (resp. $s'\in
(s-\epsilon,s)$) then $\rho(s)$ and $\rho(s')$ can be joined with
a broken lightlike geodesic as in Proposition \ref{plightlike}
with only one break.
\end{lemma}

\noindent {\em Proof.} (Reasoning just for the case $s'>s$). Let
$U$ be a convex neighborhood of $p=\rho(s)$. It is known that
there exists a globally hyperbolic neighborhood $\tilde{U}\ni p$,
$\tilde{U}\subset U$ which is causally convex in $U$ (i.e., such
that any causal curve in $U$ with endpoints in $\tilde{U}$ is
entirely contained in $\tilde{U}$), see \cite{MS}. Notice that
$E^+(p,\tilde{U})=\partial J^+(p,\tilde{U})$. Let $\epsilon
>0$ such that $\rho([s,s+\epsilon])\subset \tilde{U}$. For any $s'\in
(s,s+\epsilon]$, any past-directed lightlike geodesic $\beta$
starting at $p'=\rho(s')$ must cross $E^+(p,\tilde{U})$ at some
point $q$ (recall that $\beta$ cannot remain imprisoned in the
compact set $J^+(p,\tilde{U})\cap J^-(p',\tilde{U})$; notice also
that, eventually, $q=p'$ or $q=p$ may hold if $\rho$ is
lightlike). Thus, the unique (up to reparametrization) broken
lightlike geodesic $\gamma$ in $\tilde{U}$ which goes from $p$ to
$q$ and then to $p'$ is the required one.

Even more, in the case dim$\,\M\geq 3$ and $X$ geodesic, there are
at most two such broken geodesics $\gamma_1, \gamma_2$ which
connect $p, p'$ and are integral curves of $X$ at some point (if
they existed, one of them $\gamma_1$ would be obtained by taking
$\beta$ as the integral curve of $X$ through $p'$, and the other
one $\gamma_2$, analogously starting with an integral curve from
$p$). Thus, it is enough to construct $\gamma$ by choosing $\beta$
in a direction different to the velocities of $\gamma_1$ and $
\gamma_2$ on $p'$. The remainder for the case $\rho$ timelike is
straightforward. \cvd

\smallskip

\noindent {\em Proof of Proposition \ref{plightlike}}. Each
interval $[s_i,s_{i+1}]$ can be covered by open subsets type
$(s-\epsilon, s+ \epsilon)$, $(s_{i+1}-\epsilon, s_{i+1}]$,
$[s_{i}, s_{i}+\epsilon )$, with $\epsilon$ satisfying the
properties of Lemma \ref{llightlike}. Now, choose $\delta$ small
enough to make each $(s-\delta, s+\delta) \cap [s_i,s_{i+1}]$
included in one of these open subsets (i.e., $\delta$ is taken
smaller than a Lebesgue number of the covering) with $s_{i+1}=s_i
+ k_i \delta$ for some positive integer $k_i$. Then, the result
follows by joining each $\rho(s_{i}+k\delta ), \rho(s_i + (k+1)
\delta )$, (for $k=0,1,\dots , k_i-1$ and all $i$), as in Lemma
\ref{llightlike}. \cvd

In the case of Mp-waves, broken lightlike geodesics as in
Proposition \ref{plightlike} for $X=\partial_v$ will be  chosen.
Summing up, the following result (and its analog for the future
case) will be used  systematically. 

\begin{corollary}\label{cLIGHT} Let $\M$ be a strongly causal  Mp-wave and $P$ be a TIP.
Then, there exists an inextendible future-directed lightlike curve
$\gamma$ (in fact, a broken  geodesic  without conjugate points)
at no point proportional to $\partial_v$, such that $P=
I^-[\gamma]$ and $\uparrow P= \uparrow \gamma$.

Conversely, if $\gamma$ is any inextendible future-directed causal
curve with $\gamma \subset I^-[\gamma]$ then $P= I^-[\gamma]$ is a
TIP and $\uparrow P= \uparrow \gamma$.
\end{corollary}

\section{Fermat's arrival function and functional approach} \label{s4}

Vector field $\partial_v$ allows to define an ``arrival function''
analogous to classical Fermat's time arrival one, as well as an
associated functional. In order to carry out the analogy, consider
first  the simple case of a product spacetime\footnote{As
Causality is conformal invariant, this also corresponds to both,
the standard static case, and  the case of GRW (Generalized
Robertson-Walker spaces). Nevertheless, the construction can be
carried out in the much more general setting of splitting type
spacetimes (which include all the globally hyperbolic spacetimes
\cite{BeSa}); see \cite{Sa-Bari} for a general detailed study, or
\cite[Sect. 3]{S} for the case GRW.} $(S\times \R , g= g_S-dt^2)$,
where $(S,g_S)$ is a Riemannian manifold and $\partial_t$ points
out to the future. (Notice that, if $F\equiv 0$, a Mp-wave can be
regarded as one such product spacetime with $S=M\times \R$, after
a change of the coordinates $u,v$.) Let $x_0, x_1\in S, \Delta
>0$. For any piecewise smooth curve $y: [0,\Delta] \rightarrow S$
with endpoints $y(0)=x_0, y(\Delta ) =x_1$  a unique
future-directed lightlike curve $\gamma(t)=(y(s(t)), t), t\in
[0,T]$ can be constructed, being $s(t)$ and $T=T[y]$ determined by
$g(\dot \gamma, \dot \gamma )\equiv 0, s(0)=0, s(T)=\Delta$. So,
if $\C \equiv \C(x_0,x_1; \Delta)$ denotes the set of all such
curves $y=y(s)$, a functional
$$ \J: \C\rightarrow \R, \quad y\mapsto T[y]$$
is obtained. Now, consider the (future) {\em time arrival map}
$$ T: S\times S \rightarrow \R, \quad \quad (x_0,x_1) \mapsto  T(x_0,x_1):= {\rm Inf}_{\C}\J.$$
Easily, one has:
$$ (x_0,t_0) \ll (x_1,t_1) \quad \Leftrightarrow \quad T(x_0,x_1) < t_1-t_0 .$$
In fact, $T(x_0,x_1)$ is the (Fermat) minimum arrival time of a
future-directed lightlike curve from $(x_0,0)$ to the line
$\{x_1\}\times \R$. Notice that in this simple case function $T$
is always finite and continuous, and  essentially the same
function is obtained if past-directed causal curves are taken (see
\cite{Sa-Bari}). Next, our aim is to make a similar construction
for any Mp-wave (\ref{m-general}) but now playing $\partial_v$ the
role of $\partial_t$. The construction can be also generalized to
Eisenhart metrics \cite{Min}. Previously, observe that formulas
(\ref{gg}) and (\ref{m-general}) yield, respectively, the
following two lemmas.
\begin{lemma}\label{l1}  For any $z_{0}=(x_{0},u_{0},v_{0})\in \M$,
$I^{+}(z_{0})\subseteq M\times (u_{0},\infty)\times\R$ (resp.
$I^{-}(z_{0})\subseteq M\times (-\infty,u_{0})\times\R$).
\end{lemma}
\begin{lemma}\label{primaria} Let $z_{0}=(x_{0},u_{0},v_{0})$, $z_{1}=(x_{1},u_{1},v_{1})$, $\Delta u=u_{1}-u_{0}$. Any 
causal curve in $\M$ with endpoints $z_{0}$, $z_{1}$ and velocity
not proportional to $\partial_v$ at any point, satisfies $|\Delta
u| \neq 0$ and can be uniquely reparametrized as
$\gamma(s)=(x(s),u(s),v(s))$, $\forall s\in I=[0,|\Delta u|]$,
$\gamma(0)=z_{0}$, in such a way that ${\gamma}(s)$ satisfies:
\begin{itemize}
\item[(a)] The $u-$component is written as:
\begin{equation} \label{ejbis}
u(s)(\equiv u_\nu(s)):=u_{0}+ \nu s, \quad \quad
\forall s \in I 
\end{equation}
where $\nu = \frac{\Delta u}{|\Delta u|}$ i.e., $\nu=1$ when
$\gamma$ is future-directed and $\nu=-1$ when past-directed,
\item[(b)] putting
$E(s)=\langle\dot{\gamma}(s),\dot{\gamma}(s)\rangle_{L}$
($E(s)\leq 0$, $\forall s\in I$), then
\begin{equation}
\label{ev}
v(s)=v_{0}+ 
\frac{\nu}{2}\int_{0}^{s}(-E(\sigma)+|\dot{x}(\sigma)|^2-F(x(\sigma),u_{\nu}(\sigma)))d\sigma,
\quad \forall s\in I.\end{equation}
\end{itemize}
\end{lemma}
Now, given any $(x_0,u_0), (x_1,u_1)\in M\times \R$, put $\Delta u
= u_1-u_0$ and assume  $|\Delta u| \neq 0$. For each piecewise
smooth curve $y: [0, |\Delta u|] \rightarrow M$ with endpoints
$x_0,x_1$, consider the unique lightlike curve $z(s)=(y(s),
u_\nu(s), v_y(s)), s\in I=[0,|\Delta u|]$, $u_\nu$ as in
(\ref{ejbis}), where $v_y(s)$ is determined from (\ref{ev}) (and
thus, depends implicitly on $\nu$) by putting $E(s) \equiv 0,
v_{0}=0, x\equiv y$. So, if $\C (\equiv \C (x_0,x_1;|\Delta u|))$
denotes the set of all such curves $y$, a functional
\[
\C\rightarrow \R, \quad y\mapsto v_y(|\Delta u|)
\]
is obtained. In fact, define functional $\J_{u_0}^{\Delta u}:
\C\rightarrow \R$:
\begin{equation} \label{ej}
 \J_{u_0}^{\Delta u}(y)=
\frac{1}{2} \int_{0}^{|\Delta
 u|}(|\dot{y}(s)|^2-F(y(s),u_\nu(s)))ds.
 \end{equation}

Notice that, from the expression (\ref{ev}) for the component
$v_y(s)$ we have:
$$
v_y(|\Delta u|)=  \nu \J_{u_0}^{\Delta u}(y).
$$
Now, consider the  {\em  arrival map} $V: (M\times \R)\times
(M\times \R) \rightarrow [-\infty, \infty]$, \be \label{V}
((x_0,u_0),(x_1,u_1)) \longmapsto \;\; V((x_0,u_0),(x_1,u_1)):=
{\rm Inf}_{\C}\J_{u_0}^{\Delta u} \in [-\infty, \infty)\ee
($\Delta u =u_1-u_0$; for convenience, $V=\infty$ if $u_0=u_1$),
which satisfies the triangle inequality \be \label{triangle}
V((x_0,u_0),(x_2,u_2)) \leq V((x_0,u_0),(x_1,u_1)) +
V((x_1,u_1),(x_2,u_2)),\ee whenever $u_0<u_1<u_2$ or
$u_0>u_1>u_2$. Even more, from the expression (\ref{ej}) it
directly follows that $V$ is symmetric, i.e.: \be
\label{symmetric} V((x_0,u_0),(x_1,u_1)) = V((x_1,u_1),
(x_0,u_0)),\ee whenever $u_0\neq u_1$. From the construction, the
following result (which shows that this function plays a similar
role to time arrival Fermat's one) holds.

\begin{proposition}\label{l2} For every $z_{0}=(x_{0},u_{0},v_{0})\in \M$,
$x_{1}\in M$, $u_1 \in \R\backslash \{ u_0\}$:

If $u_{1}>u_{0}$ then $z_1= (x_{1},u_{1},v_{1})\not\in
I^{-}(z_{0})$
and: $$z_1= (x_{1},u_{1},v_{1})\in I^{+}(z_{0}) \Leftrightarrow 
v_{1}-v_0> V((x_0,u_0),(x_1,u_1)).$$

If $u_{1}<u_{0}$ then $z_1= (x_{1},u_{1},v_{1})\not\in
I^{+}(z_{0})$ and:
$$z_1= (x_{1},u_{1},v_{1})\in I^{-}(z_{0}) \Leftrightarrow
v_{1}-v_0< -V((x_0,u_0),(x_1,u_1)).$$

\end{proposition}

\noindent  {\it Proof.} Clearly, the second case follows from the
first one\footnote{In what follows, even though the results will
be stated for both, past and future, the proofs will be done only
for one of them if there is no possibility of confusion.} and,
within this case, the first assertion follows from Lemma \ref{l1}.
Then:

$(\Rightarrow )$ Consider a timelike connecting curve $\rho$, and
construct the lightlike broken geodesic
$\gamma(s)=(y(s),u(s),v(s))$ provided by Proposition
\ref{plightlike} with $X=\partial_v$.   Now, the $y(s)$ part
yields the   non-strict inequality, which is sufficient as the
equality cannot hold ($I^{+}(z_{0})$ is open and, thus, the
non-strict inequality would follow also for a smaller $v_1$).

$(\Leftarrow$) If $\Delta u >0$ then $ V((x_0,u_0),(x_1,u_1))$ is
the infimum of all the $v_y(|\Delta u|)$ for lightlike curves  (at
no point tangent to $\partial_v$) joining $(x_0,u_0,0)$ with the
line $\{(x_1,u_1)\}\times \R$. Thus, for some sequence
$\{\epsilon_m\}_m \searrow 0$, the point $z_0$ can be joined with
$p_m:= (x_1,u_1,v_0 + V((x_0,u_0),(x_1,u_1))+\epsilon_m)$ by means
of a future-directed lightlike curve,  and, for $m$ big enough,
$p_m$ can be joined with $z_1$ by means of a (future-directed)
integral curve of $\partial_v$. Thus, $z_0 < p_m < z_1$ and, as
the three points do not lie on an (unbroken) lightlike geodesic,
$z_0 \ll z_1$. \cvd

\begin{remark}\label{re-4.2}
(1) {\em Proposition \ref{l2} also ensures that Mp-waves do not
admit event horizons (according to the criterion suggested in
\cite[Sections 2.2.4]{HR2} and refined in \cite[Subsect.
3.2]{FShonnef}), as any event $z_0$ can be joined with any line
$\{(x_1,u_1)\}\times \R$ by means of either a future-directed (if
$u_0<u_1$) or a past-directed (if $u_0>u_1$) timelike curve. This
is similar to the inexistence of horizons in any spacetime which
is standard static (on the whole manifold).

(2) Due to the nature of our problem the domain of functional
$\J_{u_0}^{\Delta u}$ will be the set of {\em piecewise smooth}
curves $\C (\equiv \C (x_0,x_1;|\Delta u|))$. In fact, we will be
interested in the qualitative properties of the infimum of
$\J_{u_0}^{\Delta u}$ when $\Delta u$ tends to some
$\Delta_{\infty}$, but not in the existence of a curve minimizing
$\J_{u_0}^{\Delta u}$. When such a curve becomes relevant, a
typical technical step is to enlarge $\cal C$ by including curves
with a lower degree of smoothness ($H^1$ curves)
---remarkably, this happens in the problem of geodesic
connectedness, see \cite{CFS2}. But even in the case of
considering curves in such a enlarged ${\cal C}$, the
corresponding curves of the Mp-wave (constructed according to
Lemma \ref{primaria}) are still causal \cite[Appendix]{CFS2}, in
the continuous sense explained in Subsection \ref{s3.1}.}
\end{remark}

\begin{figure}
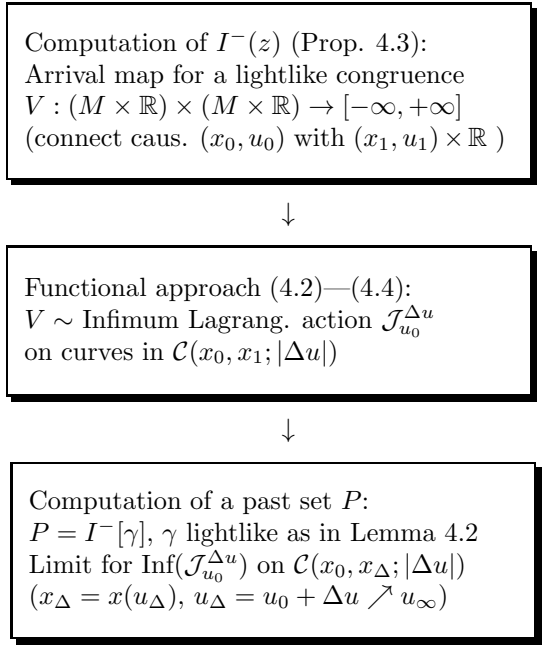
 \begin{center}
\noindent \scaja{6em}{18.5em}{
Computation of $I^{-}(z)$ (Prop. \ref{l2}): \\
Arrival map for a lightlike congruence  \\
$V:(M\times\mathbb{R})\times (M\times\mathbb{R})\rightarrow
[-\infty,+\infty]$
\\
(connect caus. $(x_0,u_0)$ with $(x_1,u_1)\times\mathbb{R}$ ) }
\par

\caja{3em}{2em}{\mbox{ }\hspace*{4em}$\downarrow$}
\par
\noindent \scaja{5em}{18.5em}{
Functional approach (\ref{ev})---(\ref{V}): \\
$V\sim$ Infimum Lagrang. action ${\cal J}_{u_0}^{\Delta u}$\\
on curves in ${\cal C}(x_0,x_1;\vert \Delta u\vert)$ }
\par

\caja{3em}{2em}{\mbox{ }\hspace*{4em}$\downarrow$}
\par
\noindent \scaja{6em}{18.5em}{
Computation of a past set $P$: \\
$P=I^{-}[\gamma]$, $\gamma$ lightlike as in Lemma \ref{primaria} \\
Limit for Inf(${\cal J}_{u_0}^{\Delta u}$) on ${\cal C}(x_0,x_{\Delta};\vert \Delta u\vert)$ \\
($x_{\Delta} = x(u_{\Delta})$, $u_{\Delta} = u_0+\Delta u\nearrow
u_{\infty}$) } \end{center}
\caption{Emergency of the functional approach
}

\end{figure}

\section{Conditions on function $F$ and functional
${\cal J}$} \label{s5}

In order to get more information about the causal cones of these
spacetimes, some technical conditions on functional ${\cal
J}_{u_{0}}^{\Delta u}$ become crucial. These conditions are
satisfied under natural restrictions on the growth of $F$. So, let
us define first such relevant types of growth.

\begin{definition}\label{superqu} Let $M$ be a connected Riemannian manifold,
and consider the chosen point $\overline{x}\in M$ in (\ref{dist}).
A function $F:M\times\R\rightarrow\R$ will be said:
\begin{itemize}
\item[(i)] {\em superquadratic}   if $M$ is unbounded and contains
a sequence of points $\{p_{m}\}_{m}\subset M$ such that
$|p_{m}|\rightarrow\infty$  and
\[
 R_{1}\cdot |p_{m}|^{2+\epsilon}+R_{0}
 \leq F(p_{m},u)
 \qquad \forall
u\in\R,
\]
for some $\epsilon,R_{1},R_{0}\in\R$ with $\epsilon,R_{1}>0$.
\item[(ii)] {\em (spatially) at most quadratic} if there exist
continuous functions $R_{0}(u), R_{1}(u)>0$ such that \be
\label{atmost} F(x,u)\leq R_{1}(u)|x|^{2}+R_{0}(u)\qquad \forall
(x,u)\in M\times\R. \ee
 Even more: (a) if (\ref{atmost}) holds when
$|x|^{2}$ is replaced by $|x|^{2-\epsilon(u)}$ for some continuous
$\epsilon(u)>0$, function $F$ is called {\em (spatially)
subquadratic}, and (b) if $M$ is unbounded and a lower bound
analogous to (\ref{atmost}) also holds, i.e.,
$$
R^-_{1}(u)|x|^{2}+ R^-_{0}(u) \leq F(x,u)\leq
R_{1}(u)|x|^{2}+R_{0}(u)
$$
$R^-_1(u)>0$ then $F$ is {\em (spatially) asymptotically
quadratic}.
 \item[(iii)] {\em $\lambda$-asymptotically
quadratic (on Mp-causal curves)}, with $\lambda >0$, if $M$ is
unbounded and there exist continuous functions
$R_{0}(u),R_{1}(u)>0$ and a constant $R_{0}^{-}\in\R$ such that:
\[
\frac{\lambda^{2}|x|^{2}+R^{-}_{0}}{u^{2}+1}\leq F(x,u)\leq
R_{1}(u)|x|^{2}+R_{0}(u)\quad \forall (x,u)\in M\times\R.
\]
\end{itemize}
\end{definition}

\begin{remark} \label{r-ConditF} {\em
(1) Of course, these definitions are independent of the choice of
$\overline{x}\in M$ in (\ref{dist}). The exact value of functions
$R_0, R_1$ is not relevant for the definitions  and, thus, no more
generality is gained if, say, a term in $|x|^{2-\epsilon(u)}$ is
added to the right hand side of the inequalities in {\em (ii)} and
{\em (iii)}. Obviously:
\begin{center}
subquadratic $\Rightarrow$
 at most quadratic $\Rightarrow$ no superquadratic

 $\lambda$-asymptotically quad. $\Rightarrow$ asymptotically
quad. $\Rightarrow$ at most quad.
\end{center}

(2) For definitions {\it (ii)} the possible growth of $F$ with $u$
is essentially irrelevant (as $R_{i}$, $R^-_{i}$ depend
arbitrarily on $u$). Nevertheless, this is not the case for the
lower bound ($\leq$) in {\it (iii)}. The reason is that now the
minimum quadratic behavior on $F$ is required when computed on
causal curves, i.e. for functions type $u\mapsto F(x(u),u)$. If
$\lambda$, $R_{0}^{-}$ depended arbitrarily on $u$, the inequality
would be very weak. In principle,  one would be forced to make the
bound independent of $u$, i.e., type
$\lambda^{2}|x|^{2}+R^{-}_{0}$. Nevertheless, we  allow a
weakening of this bound just rescaling $|x|$ by dividing it by the
same power of $u$, and even weaker conditions (as (\ref{sturm})
below) would suffice.

(3) Notice  that conditions {\it (i)}, {\it (ii)(b)} and {\it
(iii)} impose restrictions on the minimal growth of $F$ for large
$x$ and, thus, $M$ is required to be unbounded. Nevertheless,
condition (\ref{atmost}) and {\it (ii)(a)} only bounds the upper
growth of $F$ and, so, if $M$ is a bounded manifold, these
definitions also make sense. In particular, any function $F$ on a
compact $M$ will be regarded as subquadratic.

(4) As proved by the authors in \cite{FS}, if $F$ is at most
quadratic then the corresponding Mp-wave is strongly causal.
Moreover, if the Riemannian manifold $M$ is complete and $F$ is
subquadratic then the Mp-wave is globally hyperbolic. It is worth
pointing out that Hubeny, Rangamani and Ross also studied stable
causality by constructing explicitly time functions \cite{HRR},
and Minguzzi \cite[Th. 5.5]{Min} related analityc properties of
${\cal J}$ (in the more general framework of Eisenhart metrics)
with the possible causal simplicity of the spacetime.

}\end{remark} The following two technical conditions on ${\cal
J}_{u_{0}}^{\Delta u}$ will be extensively used.

\begin{definition}\label{3.3}  We will say that a Mp-wave $\M$ satisfies
hypothesis:
\begin{itemize}
\item[{\bf (H1)}.] If, for each $u_0\in \R$, there exists
$\Delta_{0} (=\Delta_0(u_0))\geq 0$ such that for every $\Delta u
> \Delta_{0}$ (resp. $\Delta u < -\Delta_{0}$), there exists a sequence of piecewise smooth loops
$x_{m}:[0,|\Delta u|]\rightarrow M$
 with the same base point
$\bar x\in M$ (i.e., $x_{m}(0)=x_{m}(|\Delta u|)=\overline{x}$)
satisfying \be \label{eh}
\begin{array}{c}
{\cal J}_{u_{0}}^{\Delta u}(x_{m})\rightarrow
-\infty\qquad\hbox{when}\;\; m\rightarrow\infty.
\end{array}
\ee

\item[{\bf (H2)}.] If hypothesis {\bf (H1)} holds with
$\Delta_0=0$ for all $u_0$.
\end{itemize}
\end{definition}

\begin{remark} \label{rh12}
{\em Obviously, hypothesis {\bf (H2)} implies {\bf (H1)}, and
there is no loss of generality assuming that the base point $\bar
x$ is equal to the point chosen in (\ref{dist}).

Condition ({\bf H1}) can be expressed in a simpler way, because if
(\ref{eh}) holds for some $\Delta u = \Delta >0$ then it also
holds for all $\Delta u >\Delta$ (construct a piecewise smooth
curve by ``stopping'' $x_m$  during an interval of length $\Delta
u - \Delta$).}
\end{remark}
In the next two lemmas, appropriate asymptotic behaviors of $F$
are proved to be sufficient for these hypotheses.

\begin{lemma}\label{l11} Hypothesis {\bf (H2)} holds if $F$ is superquadratic and $-F$ at most quadratic.
\end{lemma}
{\it Proof.} We will consider just the case $\Delta u>0$. Choose
$0<\delta<\Delta u/2$ and take a sequence $\{p_m\}_m$ as in the
definition of superquadratic. Let the sequence of curves
$x_{m}:[0,\Delta u]\rightarrow M$ be defined as juxtapositions
$x_m= \alpha_m^{-1} \star p_m \star \alpha_m$

\begin{eqnarray} \label{eloop}
x_{m}(s)=\left\{\begin{array}{ll}
\alpha_{m}(s) & \hbox{if}\; s\in [0,\delta] \\
p_{m} & \hbox{if}\; s\in [\delta,\Delta u-\delta] \\
\alpha_{m}(\Delta u - s) & \hbox{if}\; s\in [\Delta
u-\delta,\Delta u],
\end{array}\right.
\end{eqnarray} where $\alpha_{m}:[0,\delta]\rightarrow M$
is a constant speed curve joining $\overline{x}$ to $p_{m}$
with length $L_m 
\leq |p_{m}|+1$ for all $m$ (if $M$ were complete these curves
could be chosen as minimizing geodesics of speed $L_m/\delta =
|p_m|/\delta$). Clearly, the first term of ${\cal
J}_{u_{0}}^{\Delta u}(x_{m})$ in (\ref{ej}) satisfies the bound:
\begin{equation}\label{gem1}
\int_{0}^{\Delta u}|\dot{x}_{m}(s)|^2 ds = \frac{2L_m^2}{\delta}
\leq\frac{2(|p_m|+1)^{2}}{\delta}.
\end{equation}
And, from the hypotheses on $F$, the second term satisfies:
\begin{equation}\label{gem2}
\begin{array}{rl}
-\int_{0}^{\Delta u}F(x_{m}(s),u(s))ds &
 =-\int_{0}^{\delta}F(x_{m}(s),u_{0}+s)ds -\int_{\Delta u-\delta}^{\Delta u}F(x_{m}(s),u_{0}+s)ds
  \\ & \quad -\int_{\delta}^{\Delta u-\delta}F(x_{m}(s),u_{0}+s)ds
\\ & \leq 2 \delta (\tilde R_1 L_m^2 + \tilde R_0) -(\Delta u-2\delta)(R_{1}|p_m|^{2+\epsilon}+R_{0})
\\ & = -\overline{R}_{1}\,
|p_m|^{2+\epsilon}+ \mbox{(terms in $|p_m|^2$ and lower degree)},
\end{array}
\end{equation}
for some constants $R_{1}, R_{0}, \tilde{R}_{1}, \tilde{R}_{0},
\overline{R}_{1}\in \R$, with $R_{1}, \overline{R}_1
>0$. In conclusion,  by adding (\ref{gem1}) and (\ref{gem2}) and
recalling (\ref{ej}),
\[
{\cal J}_{u_{0}}^{\Delta u}(x_{m}) 
\leq\frac{1}{\delta}(|p_m|+1)^{2}-\frac{1}{2}\left(\overline{R}_{1}\,|p_m|^{2+\epsilon}-\mbox{(terms
in lower degree)}\right),
\]
which clearly converges to $-\infty$ when $m\rightarrow\infty$, as
required. \cvd

\begin{lemma}\label{l22} Hypothesis {\bf (H1)} holds if the Mp-wave satisfies any of the following conditions:
\begin{itemize}
\item[(i)] $F$ is $\lambda$-asymptotically quadratic for some
$\lambda>1/2$. \item[(ii)] $M=\R^{n}$ and $F$ is the quadratic
form
\[
F(x,u)=\sum_{ij}f_{ij}(u)x^{i}x^{j},\qquad \hbox{with}\;\;
f_{1j}\equiv f_{j1}\equiv 0\;\;\hbox{for all}\;\; j\neq 1,
\]
$$f_{11}(u)\geq\lambda^{2}/(u^{2}+1) \quad \mbox{for large $|u|$ and some}\; \lambda>1/2 .$$
In particular, this includes the case $F(x,u)= \sum_{i=1}^n
\mu_{i} (x^{i})^{2}$
 with  $\mu_{1} > 0$.
\end{itemize}
\end{lemma}
{\it Proof.} The very rough idea can be understood as follows. The
loops $x_m$ required for {\bf (H1)} will be chosen by going and
coming back from $\bar x$ to an arbitrarily far point $p_m$,
through a suitably parametrized (almost) geodesic $x_m$.
Functional ${\cal J}_{u_0}^{\Delta u}(x_m)$ will be upper bounded
essentially by \be \label{simple} \int_0^{\Delta u}\left(\dot y^2
-R_1^- y^2\right) du \ee where $y(u)(\geq 0)$ represents the
distance along $x_m$ between $\bar x $ and $x_m(u)$, and
$R^{-}_{1}(u) \gtrsim \lambda u^{-a}$ for large $u$ and $a\leq 2$.
Recall that: (a) essentially, the contribution of the integrand of
(\ref{simple}) is positive at the extremes (i.e., the base point
of the loop), and negative around the maximum of $y(u)$, and (b)
for {\bf (H1)}, one only needs to study $\Delta u>\Delta_0$, so,
one can try to find $\Delta _0$ so big that the contribution of
the negative term in (\ref{simple}) (say, with the curve staying a
big time at $p_m$) is more important than the positive one. In
fact, this is a good strategy when $a<2$ but, in order to obtain
an optimal bound when $a=2$, the relative contributions of the
negative and positive parts of (\ref{simple}) are delicate and
depend heavily on the parametrization of the curve. So, we will
consider the Euler-Lagrange equation for this functional, that is:
$$\ddot y = -R_1^-y,$$
with $y(0)=0$. This is a concave function which, under our
hypothesis, oscillates (in the sense of Sturm-Liouville theory)
and, so, satisfies $y(\Delta u)=0$ for some $\Delta u>0$. This
will yield good candidates to extremize the functional and, then,
to obtain arbitrarily large negative values for it. These  ideas
will underlie in the following formal proof.

\smallskip

 For case
{\it (i)}, let $p_{m}\in M$ be any sequence with $\{|p_m|\}_m
\rightarrow \infty$, and $\alpha_{m}:[0,1]\rightarrow M$ a
sequence of constant speed curves joining $\overline{x}$ to
$p_{m}$, whose lengths $L_{m}$ satisfy $L_m-|p_m|\searrow 0$ fast
so that \be \label{eepsilon} (0\leq) \quad (L_m s)^{2} -
|\alpha_{m}(s)|^{2} \leq \nu_0 \quad \forall s\in [0,1] \ee for
some small $\nu_0\geq 0$ (if $M$ were complete, each $\alpha_m$
would be taken as a minimizing geodesic and (\ref{eepsilon}) would
hold for $\nu_0=0$). For some $0<\epsilon < 1$ such that still
$\epsilon\lambda>1/2$, let $y_{\epsilon}(s)$ be the solution of
the problem:
\begin{equation}\label{lu}
\left\{\begin{array}{l} \ddot{y}_{\epsilon}(s)=-R_{1}^{-}(s/\epsilon+u_{0})y_{\epsilon}(s)\qquad \hbox{with}\;\; R^{-}_{1}(u)=\lambda^{2}/(u^{2}+1) \\
\dot{y}_{\epsilon}(0)=1 \\ y_{\epsilon}(0)=0.
\end{array}\right.
\end{equation}
It is known from the very beginning of Sturm-Liouville theory that
the lower bound \be \label{sturm} \limsup_{s\rightarrow \infty}
[s^2 R_{1}^{-}(s/\epsilon+u_{0})] = \epsilon^2\lambda^2 > 1/4 \ee
is the critical one for the existence of oscillatory solutions of
(\ref{lu}), see \cite[Ch. 6.3]{Ze}. Therefore, inequality
(\ref{sturm}) ensures the existence of some $\Delta^*_{0}>0$
(which may depend on $\epsilon$) such that
$y_{\epsilon}(\Delta^*_{0})=0$ (see \cite[Th. 9]{Hi} as a precise
result).

 From (\ref{lu}), obviously
\[
\dot{(\dot{y}_{\epsilon}y_{\epsilon})}=\dot{y}_{\epsilon}^{2}-R_{1}^{-}(s/\epsilon+u_{0})y_{\epsilon}^{2}
\]
and  integrating:
\begin{equation}\label{ya}
\int_{0}^{\Delta^*_{0}}\dot{y}_{\epsilon}(s)^{2}ds-\int_{0}^{\Delta^*_{0}}R_{1}^{-}(s/\epsilon+u_{0})\cdot
y_{\epsilon}(s)^{2}ds=\dot{y}_{\epsilon}(\Delta^*_{0})y_{\epsilon}(\Delta^*_{0})-\dot{y}_{\epsilon}(0)y_{\epsilon}(0)=0.
\end{equation}
Now, for the chosen $\epsilon \in (0,1)$, put
\[
\Delta u:=\Delta^*_{0}/\epsilon \qquad 
\qquad z(s):=y_{\epsilon}\left(\epsilon\cdot s\right),
\]
and notice:
\[
\begin{array}{l}
\int_{0}^{\Delta u}\dot{z}(s)^{2}ds-\int_{0}^{\Delta
u}R_{1}^{-}(s+u_{0})\cdot
z(s)^{2}ds \qquad\qquad\qquad\qquad\qquad \\
\qquad\qquad\qquad\qquad\qquad =
\epsilon\int_{0}^{\Delta^*_{0}}\dot{y}_{\epsilon}(s)^{2}ds-\frac{1}{\epsilon}\int_{0}^{\Delta^*_{0}}R_{1}^{-}(s/\epsilon
+u_{0})\cdot y_{\epsilon}(s)^{2}ds<0,
\end{array}
\]
the last inequality clearly from (\ref{ya}). In conclusion, the
sequence of curves
\[
x_{m}(s):=\alpha_{m}(z(s)/z_{max}),\qquad
z_{max}:=\hbox{max}\{z(s): s\in [0,\Delta u]\}
\]
will do the job for $\Delta u$, i.e.:
\[
\begin{array}{rl}
2{\cal J}_{u_{0}}^{\Delta u}(x_{m}) & =\int_{0}^{\Delta
u}|\dot{x}_{m}(s)|^2 ds -\int_{0}^{\Delta u} F(x_{m}(s),u(s))ds \\
& \leq \int_{0}^{\Delta u}|\dot{x}_{m}(s)|^2 ds -\int_{0}^{\Delta
u}(R_{1}^{-}(s+u_{0})|x_m(s)|^{2}+R_{0}^{-}(s+u_{0}))ds \\
& \leq \frac{L_{m}^{2}}{z_{max}^{2}}\left(\int_{0}^{\Delta
u}\dot{z}(s)^{2}ds -\int_{0}^{\Delta u}R_{1}^{-}(s+u_{0})\cdot
z(s)^{2}ds\right) \\ & \quad -\int_{0}^{\Delta
u}R_{0}^{-}(s+u_{0})ds+\nu_0\int_{0}^{\Delta u}
R_{1}^{-}(s+u_{0})ds
\\ & \rightarrow -\infty ,
\end{array}
\]
the last limit because $L_m\rightarrow \infty$ and the term in
parentheses is negative. Notice that this divergence is shown for
$\Delta u=\Delta^*_{0}/\epsilon$, which is sufficient according to
Remark {\ref{rh12}.

\smallskip

Finally, for {\it (ii)} repeat the same reasoning but taking
instead the sequence of loops $x_{m}(s)=(x^{1}_{m}(s),0,\ldots,0)$
with $x^{1}_{m}(s)=L_{m}\cdot z(s)/z_{max}$ (here $z(s)$ is
derived analogously but using the lower bound for $f_{11}$ instead
of $R_{1}^{-}$). \cvd

\begin{remark} \label{rsturm} {\em Relevant types of plane waves and pp-waves satisfy some of
the sufficient conditions in Lemmas \ref{l11}, \ref{l22}.
Moreover, the behavior of $F$ under condition {\it (i)} of Lemma
\ref{l22} is quite general and the estimates optimal.
Nevertheless, we have not tried to give a more general (but
probably less simple and transparent) result. In fact, this case
{\it (i)} does not include the case {\it (ii)}, which is
completely independent. Roughly, condition {\bf (H1)} holds when
the system corresponding to (\ref{lu}) admits two zeroes. In
particular, this happens when $F$ behaves at least
 quadratically $\sim \lambda^{2}(|x|/u)^2$, $\lambda>1/2$
 (or just satisfying (\ref{sturm})) on the $(x,u)$ part of a sequence of causal curves in
$\M$ with unbounded component $x$. So, a direction in the  $M$
part with this behaviour (where $|x|/u$ can be regarded as a sort
of ``rescaled distance'') suffices, see also Remark
\ref{r-ConditF} (2). This turns out the key behavior for the
1-dimensional character of the causal boundary.

On the other hand, by using Sturm-Liouville theory one can find
conditions subtler than ``$\lambda$-asymptotically quadratic with
$\lambda
>1/2$'' (or directly (\ref{sturm})) in order to obtain the required
oscillatory behavior for (\ref{lu}) and, thus, {\bf (H1)} (see for
example \cite[Th. 10]{Hi}, \cite[Ch. 6.3]{Ze}). Nevertheless, in
the natural types of asymptotic behaviors considered here, our
estimates (for $\lambda$, powers of the distance and dependence on
$u$) are the optimal ones, as shown in the explicit counterexample
of Subsection \ref{s9.1}.}
\end{remark}

\begin{figure}[h]

$\left. \begin{array}{c} \hbox{Superquadratic}\, F \\ + \\
\hbox{At most quadr.} \, -F
\end{array}\right\}$ $\stackrel{\hbox{Lem \ref{l11}}}{\Longrightarrow}
(H2) \sim \left(\begin{array}{c} I^+(z_0) \, \hbox{contains}  \\
\hbox{region}\, u>u_0
\end{array}\right) \stackrel{\hbox{Th. \ref{primera}}}{\Longrightarrow} \hbox{Non-distinguishing} $

\vspace*{3mm}

$\left. \begin{array}{c} \begin{array}{c} \lambda -\hbox{Asymp
quad.} \\
\hbox{with}\, \lambda>1/2 
 \end{array} \\ \hbox{ or}\\
 \begin{array}{c} \hbox{analogous condit.} \\ \hbox{ in some direction  } \end{array}\\
\hbox{or}\\ \begin{array}{c} \hbox{weaker Sturm }\\
\hbox{condit. as (\ref{sturm}) }\end{array}
\end{array}\right\}$ $\stackrel{\hbox{Lem \ref{l22}}}{\Longrightarrow}
(H1) \sim \left(\begin{array}{c} I^+(z_0) \, \hbox{contains}  \\
\hbox{reg.}\,
 u>u_0+\Delta_0
\end{array}\right) \stackrel{\hbox{Fig. \ref{f4}}}{\Longrightarrow} \left(\begin{array}{c}  P, \uparrow P\; \hbox{explicit}\\ \partial \M\; \hbox{low dim}
\end{array}\right)
$

\begin{center}
\caption{Consequences of the behaviour of $F$:
 technical
conditions (H1), (H2) (Defn. \ref{3.3}) vs asymptotic conditions
(Defn. \ref{superqu}). The ($\lambda\leq 1/2$)-asymptotic case
becomes critical (Section \ref{s9.1}) and the subquadratic case
globally hyp.  with expected higher dimension of $\partial \M$ (at
least in the case $M$ complete, Sections \ref{s9.2}, \ref{s9.3}.)}
\end{center}
\end{figure}

\section{Non-distinguishing Mp-waves}\label{s6}

In this section previous results are applied in order to prove
that, when  $F$ is superquadratic, the causal structure of
Mp-waves may become ``degenerate'' in certain sense. More
precisely, such Mp-waves will not be dis\-tin\-gui\-shing. As this
is the minimum hypothesis in order to identify the points of $\M$
with pairs $(P,F)$, these $M$p-waves cannot admit a causal
boundary. Nevertheless, this does not mean that these spacetimes
may not be useful from the AdS/CFT viewpoint
\cite{HRRb}\footnote{Figure 1 in this reference may also help to
understand the geometric situation.}.

\subsection{The general result}\label{s6.1}

\begin{theorem}\label{primera} A Mp-wave satisfying condition {\bf
(H2)} is neither future nor past-dis\-tin\-gui\-shing. More
concretely, under this hypothesis
\[
\begin{array}{c}
I^{+}(z_{0})=M\times (u_{0},\infty)\times\R\qquad\forall z_{0}\in \M \\
 I^{-}(z_{0})=M\times (-\infty,u_{0})\times\R \qquad \forall
z_{0}\in \M.
\end{array}
\]
In particular, this happens if $F$ is  superquadratic and $-F$ at
most quadratic.
\end{theorem}
{\it Proof.} From Lemma \ref{l1}, to show
\[
M\times (u_{0},\infty)\times\R \subseteq I^{+}(z_{0})
\]
 suffices, and by Proposition \ref{l2}, it is enough to check
\be \label{eee} {\rm Inf}_{\C}\J (=V((x_0,u_0),(x_1,u_1))) =
-\infty \quad\hbox{when}\;\; u_{1}>u_{0}.
 \ee
Thus, put $\Delta u = u_1-u_0$ and choose $0<\delta<\Delta u/2$.
From ${\bf (H2)}$ there exists a sequence $x_m:[\delta, \Delta
u-\delta]\rightarrow M$ satisfying the corresponding divergence
(\ref{eh}). So, if $\alpha :[0,\delta]\rightarrow M$ and $\beta
:[\Delta u-\delta,\Delta u]\rightarrow M$ are two fixed smooth
curves joining $x_{0}$ to $\overline{x}$ and $\overline{x}$ to
$x_{1}$, respectively, the sequence of juxtaposed curves (as in
(\ref{eloop})) $\{\beta \star x_m \star \alpha\}_m$ satisfies the
required divergence for (\ref{eee}). \cvd

\begin{remark} {\em If $F$ is lower bounded then $-F$ is at most
quadratic trivially. Thus, Th. \ref{primera} extends our previous
result \cite[Prop. 2.1]{FS}. On the other hand, Th. \ref{primera}
can be extended clearly to obtain the cases future and past
distinguishing independently (split  condition {\bf (H2)} in
future and past cases in an obvious way). }\end{remark}

As it is well-known, plane waves are always strongly causal, and
thus, cannot lie under the hypotheses of previous theorem.
However, this result is useful to decide if many other pp-waves of
possible interest to string theorists can admit a causal boundary.

\subsection{Some remarkable examples}\label{s6.2}

Essentially, the following examples are taken from Hubeny and
Rangamani \cite{HR}. The expectations to obtain a 1-dimensional
boundary  are truncated here, as the pp-waves may be
non-distinguishing --a possibility already suggested by the own
authors and Ross in \cite{HRR}.

\smallskip

\noi {\bf (1)} Consider the pp-wave $\M=\R^{n}\times \R^{2}$ with
\begin{equation}\label{expression}
F(x^{1},\dots, x^n, u)= {\rm cosh} x^{1}-\cos x^{2}.
\end{equation}
This spacetime leads to the ${\cal N}=2$ sine-Gordon theory on the
world-sheet in light-cone quantization.  $\M$ does not admit a
causal boundary, since function $F$ in (\ref{expression}) is
bounded below and superquadratic (take for example
$p_{m}=(m,0,\ldots,0)$ in Definition \ref{superqu} {\em (i)}), and
so, Theorem \ref{primera} (or previous computations in
\cite{FS,HRR}) applies.

\smallskip

\noindent {\bf (2)} Consider the  generalization of previous case
to a pp-wave with
\[
F(x^{i},u)= \sum_{j}f_{j}(x^{j}).
\]
In \cite{HR} the authors studied the case of a single coordinate
$F(x,u)=f(x)$. They stated that the causal boundary is
1-dimensional whenever $f(x)$ is bounded from below and, in
addition, $f(x\rightarrow\pm\infty)\rightarrow +\infty$. This
agrees with our results if $f \sim x^2$ at infinity. However, from
Theorem \ref{primera}, these conditions lead to non-distinguishing
spacetimes whenever $f$ (or one of the functions $f_j$) behaves
superquadratically (for example, $F(x,u)=x^{4}$) and thus, the
boundary is not well defined.

\smallskip

\noindent {\bf (3)} Another examples  in \cite{HR} are the
4-dimensional vacuum pp-wave spacetime with $F((x^1,x^2),u) =
-\sin x^1 e^{x^2}$ or the $5$-dimensional pp-wave
$\M=\R^{3}\times\R^{2}$ with
\[
F(r,\theta,\phi,u)= r^{3}(5\cos^{3}\theta-3\cos\theta).
\]
In these cases, function $F$ is superquadratic but $-F$ is not at
most quadratic. However, condition {\bf (H2)} still holds because
both conditions on $F$ hold in at least one direction, say
$x_1=-\pi/2$ for the first example, or $\theta=0$ for the second
one (explicitly, take, say, $x_{m}(s)=(m\sin\frac{\pi}{\Delta u}
s,0,0)$ in the second example). Thus, the causal boundary is not
well defined again by Theorem \ref{primera}.

\smallskip

\noindent {\bf (4)} Finally, consider an arbitrary $4$-dimensional
vacuum pp-wave spacetime; i.e.,
\[
\M=\R^{4}, \quad 
\langle\cdot,\cdot\rangle_{L}
=d(x^{1})^{2}+d(x^{2})^{2}-F(x,u)du^{2} - 2du\, dv ,
\]
with function $F(x,u)$ spatially harmonic ($\partial^2_{x^{1}} F +
\partial^2_{x^{2}} F \equiv 0$). As pointed out in \cite{FShonnef}, in this
case there are only three possibilities\footnote{For each constant
$u$, either the harmonic function $F(\cdot, u)$ is superquadratic
or it is polynomically bounded, and thus, it becomes a polynomial
(of at most the  degree of the bound; in this case, $2$): the
result is well-known for holomorphic functions; for harmonic ones
stronger results can be seen, for example, at \cite[Lemma
4.1]{LT}.}: (i) either $F$ is superquadratic, and thus, the causal
boundary makes no sense in general\footnote{Moreover, in this case
the pp-wave would be incomplete, according to a conjecture by
Ehlers and Kundt \cite{EK}.}, or (ii)
$F(x^{1},x^{2},u)=f(u)((x^{1})^{2}-(x^{2})^{2})+2g(u)x^{1}x^{2}$,
and then we have a plane wave (see Subsection \ref{s8.3}), or
(iii) $F(x,u)=a(u)+b(u)x^{1}+c(u)x^{2}$. In this last case the
pp-wave is Lorentz-Minkowski space\footnote{Notice that the
curvature vanishes (see for example \cite[Sect. 2]{CFS},
\cite[formula (3)]{FShonnef}) and the spacetime is complete
\cite[Prop. 3.5]{CFS} and simply connected.}, and thus, the causal
boundary is the classical double cone (also for the new concept of
causal boundary \cite[Example 10.1]{F}).

\section{Boundaries in strongly causal Mp-waves} \label{s7}

From now on, the ambient hypothesis on  $\M$ will be strong
causality, so that $\M$ admits a causal boundary with a natural
topology. Nevertheless, we will state it explicitly because most
of the computations in the next subsection are valid for any
Mp-wave.

From Corollary \ref{cLIGHT}, only (right) inextendible $\nu$-{\em
lightlike curves} \be \label{curvas} \gamma:[0,\nu \Delta_\infty
)\rightarrow \M, \quad \nu \Delta_\infty\in (0,\infty], \ee with
$\dot{\gamma}(s)$ independent of $X=\partial_{v}$ for all $s$, are
needed in order to compute the pairs $(P,F)\in
\partial\M$. Here again $\nu=\pm 1$ keeps track of the causal
orientation of $\gamma$ ($\nu=1$ for future-directed $\gamma$ and
$\nu=-1$ for past directed) and, so, $I^{-\nu}[\gamma]$, $\uparrow
^\nu \gamma$ will denote $I^{-}[\gamma]$, $\uparrow  \gamma$
(resp. $I^{+}[\gamma]$, $\downarrow  \gamma$) if $\nu = 1$ (resp.
$\nu=-1$); for simplicity, the reader can consider the case
$\nu=1$ and check the final expressions for $\nu=-1$. {\em In what
follows, we will work under a reparametrization
$\gamma(s)=(x(s),u_\nu(s), v(s))$ as in Lemma \ref{primaria};
notice that $v(s)$ is given by (\ref{ev}) with $E(s)\equiv 0$.}

We also put $ \gamma(0)= z_0 = (x_{0},u_{0},v_{0})$ and $u_\infty
= u_0 + \Delta_\infty \in [-\infty, \infty]$. For any  $\nu \Delta
\in (0, \nu \Delta_\infty)$, we will consider the
restriction $
\gamma|_{[0, \nu \Delta]}$ and put $\gamma(\nu \Delta )= z_\Delta
=(x_\Delta, u_\Delta, v_\Delta)$.

\begin{remark}\label{rinextend}
{\em Recall that the curve $\gamma$ must be inextendible. As
$\gamma$ is reconstructed from its spatial part, $x$ will be said
{\em inextendible} when: (i) $\nu \Delta_\infty =\infty$ (i.e.,
$\gamma$ is inextendible in $u$), (ii) $\nu \Delta_\infty <
\infty$ but $x$ is not continuously extendible to $\nu
\Delta_\infty$ ($\gamma$ is inextendible in $x$), or (iii) $\nu
\Delta_\infty < \infty$, $x$ is continuously extendible to $\nu
\Delta_\infty$ but the {\em (total kinetic) energy} diverges,
i.e.: $(1/2) \int_{0}^{\nu\Delta_{\infty}}|\dot x(s)|^2 ds =
\infty$ ($\gamma$ is inextendible in $v$).}
\end{remark}

\subsection{General expressions for $P, F$, $\uparrow P, \downarrow F$}\label{s7.1} 

Let us start with $I^{-\nu}[\gamma]$. From Proposition \ref{l2}, a
point $\overline{z}_0=
(\overline{x}_{0},\overline{u}_{0},\overline{v}_{0})\in \M$ with
$ \nu \overline{u}_{0}<  \nu u_{\infty}$ 
lies  in  $I^{-\nu}[\gamma]$  if and only if (recall the symmetry
of $V$, see (\ref{symmetric})),
\begin{equation}\label{c}
\nu (v_\Delta - \overline{v}_{0}) > V((\bar x_0, \bar u_0),
(x_\Delta, u_\Delta ))
\end{equation}
for some $ \nu \Delta >0$ (close to $\nu \Delta_\infty$). Put \be
\label{uves}
\begin{array}{rl}
V_\Delta = & \nu (v_\Delta - v_0), \\
 V_\Delta(\bar x_0, \bar u_0)= &
V((\bar x_0, \bar u_0), (x_\Delta, u_\Delta )) (= V((x_\Delta,
u_\Delta ), (\bar x_0, \bar u_0))), \end{array}\ee that is,
\begin{equation}\label{nueva}
\begin{array}{rl}
V_\Delta = & \frac{1}{2}\int_{0}^{|\Delta|}(|\dot{x}(s)|^2
-F(x(s),u_\nu(s)))ds \\
V_\Delta(\bar x_0, \bar u_0)= & \hbox{inf}_{{\cal C}}{\cal
J}_{\overline{u}_{0}}^{\overline{\Delta}}=\hbox{inf}_{y\in {\cal
C}} \left\{
\frac{1}{2}\int_{0}^{|\overline{\Delta}|}(|\dot{y}(s)|^2
-F(y(s),\overline{u}_{0}+ \bar\nu s)) ds\right\} ,
\end{array}
\end{equation}
where
\begin{equation}\label{incr}
\overline{\Delta}:= u_\Delta -\overline{u}_{0}
\end{equation}
(here ${\cal C}\equiv {\cal C}(\bar x_0, x_\Delta;
|\overline{\Delta}|)$ is the set of piecewise smooth curves
defined in $[0,|\overline{\Delta}|]$ joining $\overline{x}_{0}$
with $x_\Delta$, and $\overline{\nu}=
\overline{\Delta}/|\overline{\Delta}|$; recall also that, by
hypothesis on $\bar z_0$, $\nu = \bar \nu$ for $\Delta$ close to
$\Delta_\infty$). Now condition (\ref{c}) translates into
\begin{equation}\label{chu}
V_\Delta- V_\Delta(\bar x_0, \bar u_0) > \nu(\overline{v}_{0} -
v_0) \quad\;\hbox{for }\; \Delta\;\hbox{close to} \;
\Delta_\infty.
\end{equation}
\begin{lemma}\label{tt}
The left-hand side of  (\ref{chu}) is non-decreasing when $\nu
\Delta \nearrow \nu \Delta_\infty$.
\end{lemma}
{\em Proof}. Close to $\nu \Delta_\infty$ and for small $\nu
\epsilon
>0$,
we have $\nu \bar u_0 < \nu u_\Delta < \nu u_{\Delta + \epsilon}
(< \nu u_\infty)$, and by using the triangle inequality
(\ref{triangle}):
\[
\begin{array}{rl}
V_{\Delta+ \epsilon} = & V_\Delta
+\frac{1}{2}\int_{|\Delta|}^{|\Delta + \epsilon| }(|\dot{x}(s)|^2
-F(x(s),u_\nu(s)))ds \\
V_{ \Delta+ \epsilon}(\bar x_0, \bar u_0) \leq  & V_\Delta(\bar
x_0, \bar u_0) +{\rm inf}_{y\in {\cal
C}'}\frac{1}{2}\int_{|\overline{\Delta}|}^{|\overline{\Delta}+
\epsilon |}(|\dot{y}(s)|^2 -F(y(s),\bar u_{0}+ \nu s))ds,
\end{array}
\]
where now ${\cal C}'$ is equal to ${\cal C}(x_\Delta,
x_{\Delta+\epsilon}; \nu \epsilon)$ up to the reparametrization
(by means of a translation) with domain $[|\overline{\Delta}|,
|\overline{\Delta} +\epsilon|]$. Thus, as claimed,
\[
V_{\Delta +\epsilon} - V_{ \Delta +\epsilon}(\bar x_0, \bar u_0)
\geq V_\Delta- V_\Delta(\bar x_0, \bar u_0) .
\]
\cvd

Thus, taking the limit  $\nu\Delta \nearrow \nu\Delta_\infty$ in
(\ref{chu}), the following result is obtained.

\begin{proposition} \label{l-I-}
Let $\gamma$ be a  inextendible  $\nu$-lightlike curve (as in
(\ref{curvas})). For each $\bar z_0=(\bar x_0,\bar u_0,\bar v_0)$,
put: \be \label{bemol-} b^-(\bar x_0,\bar u_0) = \lim_{\nu \Delta
\nearrow \nu \Delta_\infty} \left(V_\Delta- V_\Delta(\bar x_0,
\bar u_0)\right)
 \ee
(with $V_\Delta, V_\Delta(\bar x_0, \bar u_0)$ defined in
(\ref{uves})). Then:
$$
I^{-\nu}[\gamma]=\{\overline{z}_{0}\in \M: \; \nu
\overline{u}_{0}< \nu u_{\infty}\; \; \hbox{and}\; \;
b^-(\overline{x}_{0},\bar u_0)>\nu(\overline{v}_{0}-v_{0})\}. 
$$
\end{proposition}

Next, let us consider the common future (or past)
$\uparrow^\nu\gamma$ for $\gamma$. From Proposition \ref{l2}, a
point $\bar z_0=
(\overline{x}_{0},\overline{u}_{0},\overline{v}_{0})\in \M$ with
$\nu  \overline{u}_{0}\geq \nu u_\infty$ 
lies  in  $I^{+\nu}[\gamma(\nu\Delta)]$  if and only if (recall
the notation in (\ref{uves}))
\[
\nu ( \overline{v}_{0}-v_\Delta)>V_\Delta(\bar x_0, \bar u_0)
\]
that is,
\begin{equation}\label{chu-f}
V_\Delta + V_\Delta(\bar x_0, \bar
u_0)<\nu(\overline{v}_{0} - v_0). 
\end{equation}
Reasoning as in Lemma \ref{tt},  the triangle inequality
(\ref{triangle}) implies that the left-hand side of (\ref{chu-f})
is non-decreasing with $\nu \Delta$ (but now apply it taking into
account $\nu \bar u_0 > \nu u_{\Delta + \epsilon} > \nu
u_{\Delta}$, for $\nu\epsilon>0$). So, the non-strict inequality
will hold in (\ref{chu-f}) when the limit $\nu\Delta \nearrow \nu
\Delta_\infty$ is taken.
 This will be the key for the following result.

\begin{proposition} \label{l-I-f}
Let $\gamma$ be a  inextendible  $\nu$-lightlike curve (as in
(\ref{curvas})). For each $\bar z_0=(\bar x_0,\bar u_0,\bar v_0)$,
put: \be \label{bemol+} b^+(\bar x_0,\bar u_0) = \lim_{\nu \Delta
\nearrow \nu \Delta_\infty} \left(V_\Delta + V_\Delta(\bar x_0,
\bar u_0)\right)
 \ee
(with $V_\Delta, V_\Delta(\bar x_0, \bar u_0)$ defined in
(\ref{uves})). Then:

$$
\uparrow^\nu \gamma =I^{+\nu}[\{\overline{z}_{0}\in \M: \; \nu
\overline{u}_{0} \geq \nu u_{\infty}\; \; \hbox{and}\; \;
b^+(\overline{x}_{0}, \bar u_0)\leq \nu(\overline{v}_{0}-v_{0})\}] 
$$

\end{proposition}

\noindent {\em Proof.} (For $\nu=1$). The inclusion $\subseteq$
for $\uparrow \gamma $ follows easily from the reasoning above.

For the converse, let $\overline{z}'_{0}\gg \overline{z}_{0}$,
with $\overline{z}_{0}$ such that $\overline{u}_{0}\geq
u_{\infty}$ and $b^+ (\overline{x}_{0}, \bar u_0) \leq
\overline{v}_{0}-v_{0}$. We can choose
$\overline{z}'_{0}\gg\overline{z}''_{0}\gg\overline{z}_{0}$ and we
only need to show $\overline{z}''_{0}\gg \gamma(\Delta)$ for all
$\Delta$. Since $V_\Delta + V_\Delta(\bar x_0, \bar u_0)$ is
non-decreasing, the condition on $b^+(\overline{x}_{0}, \bar u_0)$
implies
\begin{equation}\label{ññ}
V_\Delta + V_\Delta(\bar x_0, \bar u_0) \leq \overline{v}_{0} -
v_0,\qquad\hbox{for all}\;\; \Delta.
\end{equation}
On the other hand, condition $\overline{z}''_{0}\gg
\overline{z}_{0}$ implies
\begin{equation}\label{ññ'}
V((\bar x_0, \bar u_0), (\bar x''_0, \bar u''_0)) < \bar
v_0''-\bar v_0 .
\end{equation}
Thus, adding (\ref{ññ}), (\ref{ññ'}) and using the triangle
inequality (\ref{triangle}):
\[
V_\Delta+ V_\Delta(\bar x''_0, \bar u''_0) < \bar v_0''- v_0 ,
\]
that is,
\[
 V_\Delta(\bar x''_0, \bar u''_0) < \bar v_0''- v_\Delta ,
\]
as required. \cvd

 Recall that, by using Lemma \ref{primaria}, the
lightlike curve $\gamma$ in previous two propositions can be
reconstructed from its initial point $\gamma(0)$, its $x$-part and
its future or past causal character $\nu=\pm 1$; in particular,
functions $b^\pm$ can be constructed from $u_0, \nu$ and curve
$x(s)$. Nevertheless, in order to obtain the sets
$\uparrow^{\nu}I^{-\nu}[\gamma]$ associated to each
$I^{-\nu}[\gamma]$ by means of these propositions, one must take
into account that technicalities appear when $\gamma$ (necessarily
a lightlike pregeodesic) is not included in $I^{-\nu}[\gamma]$ (in
fact, here perhaps
$\uparrow^{\nu}\gamma\neq\uparrow^{\nu}I^{-\nu}[\gamma]$; recall
Remark \ref{r-contraej} and Corollary \ref{cLIGHT}). Fortunately,
the following lemma shows that this situation cannot happen in our
case.
\begin{lemma} \label{l-technicality}
Let $\gamma: [0, |\Delta_\infty|) \rightarrow \M$ be a
inextendible $\nu$-lightlike curve constructed from Lemma
\ref{primaria}. Then there exists an inextendible $\nu$-timelike
curve $\rho: [0, |\Delta_\infty|) \rightarrow \M$ such that
$I^{-\nu}[\gamma] = I^{-\nu}[\rho]$ and $\uparrow^{\nu} \gamma =
\uparrow^{\nu} \rho$.
\end{lemma}
{\em Proof.} Construct  $\rho$  from $\gamma$ as follows. Take
some negative function $E(s)<0$ with
$-\int_{0}^{|\Delta_{\infty}|}E(s)ds= \epsilon\in (0,\infty)$.
Then, $\rho$ will have the same parts $u(s), x(s)$ of $\gamma$,
but compute the $v(s)$ part from (\ref{ev}) using the chosen
function $E(s)$ and replacing $v_0$ by $v_{0}-\nu\epsilon$.
Obviously, $I^{-\nu}[\gamma] \supseteq I^{-\nu}[\rho]$ and
$\uparrow^{\nu} \gamma \subseteq \uparrow^{\nu} \rho$. For the
converses, remake the proofs of Propositions \ref{l-I-},
\ref{l-I-f} for $\rho$, checking that the additional term in
$E(s)$ does not affect to the limits for $b^\pm$. \cvd

Summing up, this subtlety plus Propositions \ref{l-I-},
\ref{l-I-f} yields the following characterization of TIP's and
TIF's.

\begin{theorem} \label{t-pastsets}
Any TIP, $P$ (resp. TIF, $F$) of a strongly causal $M$p-wave
(\ref{m-general}) is constructed as follows. Take $(u_0, v_0)\in
\R^2$, a piecewise smooth curve $x: [0, |\Delta_\infty|)
\rightarrow M$ inextendible to $|\Delta_\infty|$ (in the sense of
Remark \ref{rinextend}) and the function $b^-$ associated to $u_0,
x$ and $\nu=1$ (resp. $\nu=-1$) from Lemma \ref{primaria} and
(\ref{bemol-}). Putting $\Delta_\infty = \nu |\Delta_\infty|$ and
$u_\infty = u_0 + \Delta_\infty$ one has:

$$
P=\{\overline{z}_{0}\in \M: \;  \overline{u}_{0}< u_{\infty}\; \;
\hbox{and}\; \; b^-(\overline{x}_{0},\bar
u_0)>\overline{v}_{0}-v_{0}\}
$$
$$
(\hbox{resp.} \; F=\{\overline{z}_{0}\in \M: \;  \overline{u}_{0}>
u_{\infty}\; \; \hbox{and}\; \; b^-(\overline{x}_{0},\bar u_0)>
v_{0}-\overline{v}_{0}\}).
$$
Even more, taking also the function $b^+$ from (\ref{bemol+}):
$$
\uparrow P= I^+[\{\overline{z}_{0}\in \M: \;  \overline{u}_{0}\geq
u_{\infty}\; \; \hbox{and}\; \; b^+(\overline{x}_{0},\bar u_0)\leq
\overline{v}_{0}-v_{0}\}
$$
$$
(\hbox{resp.} \; \downarrow F= I^-[\{\overline{z}_{0}\in \M: \;
\overline{u}_{0} \leq u_{\infty}\; \; \hbox{and}\; \;
b^+(\overline{x}_{0},\bar u_0)\leq  v_{0}-\overline{v}_{0}\}).
$$
\end{theorem}

{\em Proof}. Let $P$ be a TIP. By Corollary \ref{cLIGHT}, $P$ can
be written as the chronological past of a lightlike curve $\gamma$
as in (\ref{curvas}). Applying Propositions \ref{l-I-},
\ref{l-I-f} to $\gamma$ the required expressions for $P, \uparrow
P$ holds.

Conversely, let $P$ be a set defined as in the expression above,
and take the associated inextendible future-directed lightlike
curve $\gamma$ such that $P=I^-[\gamma]$. By Lemma
\ref{l-technicality}, $P$ is a TIP and $\uparrow P = \uparrow
\gamma$, as required.
 \cvd

\begin{figure}
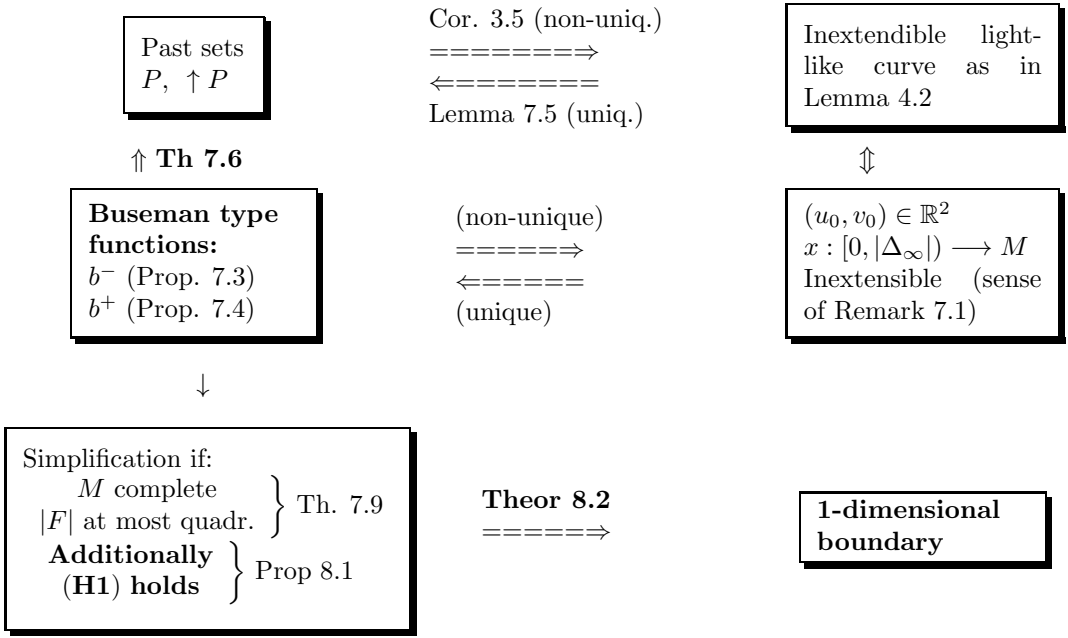


\begin{center}

\begin{tabular}{ccc}
\scaja{3em}{4em}{Past sets \\ $P$, \ $\uparrow P$} &
\caja{4em}{10em}{
    Cor. \ref{cLIGHT} (non-uniq.) \\
    =======$\Rightarrow$ \\
    $\Leftarrow$=======\\
    Lemma \ref{l-technicality} (uniq.)}
&
\scaja{4em}{9em}{Inextendible lightlike curve as in Lemma \ref{primaria}} \\
\caja{3em}{5em}{$\Uparrow$ {\bf Th \ref{t-pastsets}}} & &
\caja{2em}{5em}{$\Updownarrow$ }  \\
\scaja{5em}{8em}{{\bf Buseman type}\\ {\bf functions:} \\
$b^{-}$ (Prop. \ref{l-I-}) \\
$b^{+}$ (Prop. \ref{l-I-f}) }
 &
\caja{4em}{8em}{(non-unique)\\
=====$\Rightarrow$ \\
$\Leftarrow$===== \\
(unique)} & \scaja{5em}{9em}{
$(u_0,v_0)\in\mathbb{R}^2$\\
$x:[0,\vert \Delta_{\infty}\vert)\longrightarrow M$\\
Inextensible (sense of Remark \ref{rinextend})
} \\
\caja{4em}{8em}{\hspace{4em}$\downarrow$} & & \\
\scaja{7em}{14em}{Simplification if: \\
$\left.\begin{array}{c} M\ \textrm{complete} \\
\vert F\vert \ \textrm{at most quadr.}
\end{array}\right\}$
Th. \ref{3} \\ \vspace{0.2em} $\left. \begin{array}{c}
\textrm{{\bf Additionally}} \\ {\bf (H1)}\ \textrm{{\bf holds}}
\end{array}
\right\}$ Prop \ref{propo!!} } & \caja{3em}{6em}{ {\bf Theor 8.2} \\
=====$\Rightarrow$
\\
} & \scaja{2em}{8em}{{\bf 1-dimensional boundary} }
\end{tabular}
\end{center}

\caption{Computation of $P$, $\uparrow P$ in strongly causal
Mp-waves. The scheme of general computation in terms of Buseman
type  functions is summarized in the four upper boxes. In the two
bottom ones, under some mild technical simplifications, hypothesis
{\bf (H1)} implies the 1-dimensional boundary. In boldface crucial
conclusions, beyond the technical development.}\label{f4}
\end{figure}

\subsection{Boundary for $M$ complete, $|F|$ at most quadratic}\label{s7.2}

In order to get a manageable causal boundary we need to impose not
only strong causality  but also  a pair of  (natural and not too
restrictive) hypotheses more. The first one is the completeness of
the Riemannian part $M$. Otherwise, the Riemannian Cauchy boundary
of $M$ (i.e., the boundary for the completion as a metric space by
using Cauchy sequences) can complicate the causal boundary.
Technically, completeness yields the following well-known
property, to be used later. If the curve $x: [0, |\Delta_\infty|)
\rightarrow M, |\Delta_\infty|<\infty$ is not continuously
extendible to $|\Delta_\infty|$ (i.e., it lies in the case (ii) of
Remark \ref{rinextend}) then the completeness of $M$ implies that
its length diverges and, by Cauchy-Schwarz inequality, so does its
energy, i.e.: \be \label{ediverging}
 \int_0^{|\Delta_\infty|}|\dot x(s)|^2
 ds = \infty. \ee The second one is that not only $F$ must be at
most quadratic (which is the natural sufficient bound for strong
causality \cite[Th. 3.1]{FS}), but also so must be $|F|$.
Otherwise, other interesting geometric properties of the
spacetime, as the geodesic completeness of the whole Mp-wave, may
be destroyed (even in the simplest case of $M$ complete and $F$
independent of $u$), see Remark \ref{rpalo}.

Under these two hypotheses, we will obtain a technical property
for $\uparrow \gamma$ or $\downarrow \gamma$ (Lemma
\ref{previo}(ii)), plus a remarkable simplification for TIP's and
TIF's, namely, {\em any TIP determined by a $(\nu = 1)$-lightlike
curve with $|\Delta_\infty |<\infty$, is just the region}
$u<u_\infty$ (Lemma \ref{previo}(i)).

\begin{remark} \label{rpalo} {\rm This simplification is also pointed
out in \cite[Sect. 5.1]{HR}. Nevertheless, the hypothesis $|F|$ at
most quadratic is missing there. The following example shows that
it cannot be dropped. Consider the ($3$-dimensional) pp-wave $\M =
\R^3$ with $F(x,u)=-x^{4}$. This is globally hyperbolic (as $F$ is
subquadratic) and incomplete. In fact, the future-directed
lightlike geodesic $\gamma:[0,u_\infty)\rightarrow \M$,
$\gamma(s)=(y(s),u(s),v(s))$ determined by
$$ \quad s(y)=\int_0^y \frac{1}{\sqrt{1+\bar y^4}}d\bar y , \quad \quad  u(s)=s, \quad v(0)=0 ,$$ is
incomplete, as $u_\infty$ (the integral until $y=\infty$) is
finite. Obviously, $I^{-}[\gamma]\subset \{\overline{z}_{0}:
\overline{u}_{0}<u_\infty \}$ {\em but the inclusion is strict}.
In fact, $y(s)$  strictly minimizes functional\footnote{This can
be checked from general arguments: (a) looking at the functional
as a Lagrangian with negative ``potential'' $V=-x^4/2$, a minimum
must be attained for each $s_0\in (0,u_\infty )$, (b) this minimum
must be attained by a solution of the corresponding Euler-Lagrange
equation $\ddot x -2 x^3=0$ (one of its solutions being $y(s)$),
and (c) the boundary conditions $x(0)=y(0)=0, x(s_0)= y(s_0)$,
determine univocally the solution (recall that, from standard
theory of equations, a second solution $x(s)$ would be fixed
univocally by $x(0), \dot x(0)$, but, for example, $\dot x(0)
>\dot y(0) \Rightarrow x(s_0)
> y(s_0)$).} $\int_{0}^{s_0}(\dot{x}(s)^2 +
x(s)^{4})ds$ for all $s_0\in [0,u_\infty )$. From (\ref{V})
 the arrival function $V$ satisfies:
$$
V((0,0), (y(u), u))= \frac{1}{2} \int_{0}^{u}(\dot{y}(\sigma)^2 +
y(\sigma)^{4})d\sigma = v(u), \quad \forall u\in (0,u_{\infty}).
$$
So, from the interpretation of $V$ (Prop. \ref{l2}),  $(0, 0, v_0)
\not\in I^-(\gamma(u))$ for any $u\in (0,u_{\infty})$,  whenever
$v_0\geq 0$. }
\end{remark}

In order to avoid these difficulties, from now on we will assume
that $|F|$ is at most quadratic and $M$ complete:
\begin{lemma}\label{previo} Under these two hypotheses, if
the inextendible causal  curve $\gamma: [0,\nu
\Delta_\infty)\rightarrow \M$ satisfies $\nu \Delta_\infty <
\infty$ then:

(i) $b^-(\bar x_0, \bar u_0)= \infty$ if $\nu \bar u_0 <  \nu
u_\infty$,

(ii)  there exists $\nu \delta>0$ such that $b^+(\bar x_0, \bar
u_0)= \infty$ whenever $\nu u_\infty \leq \nu \bar u_0 < \nu
(u_\infty + \delta)$. Even more, $|\delta|= \infty$ if $F$ is
subquadratic.
\end{lemma}
{\it Proof.} Since $\gamma$ is inextendible, so is its component
$x$, thus, its energy  (\ref{ediverging}) diverges.
 From the at most quadratic behaviour of $|F|$, and the fact that the image of $u_\nu$ lies in a compact subset, we have, up to an additive
 constant:
\begin{equation}\label{des-def}
2 V_\Delta \geq A_\Delta :=\int_{0}^{\Delta} |\dot{x}(s)|^2
ds -R\int_{0}^{\Delta}
|x(s)|^2 ds,\quad\hbox{for some}\;\; R>0
\end{equation}
(recall the first formula in (\ref{nueva})).

{\em (i)}. As $V_\Delta(\bar x_0, \bar u_0)$ is obtained taking an
infimum in $\J_{\bar u_0}^{\overline{\Delta}}$ (recall
$\overline{\Delta} = u_\Delta-\bar u_0= \Delta + u_0- \bar u_0)$),
from Prop. \ref{l-I-} it is enough to exhibit a curve $y_\Delta
\in {\cal C}(\bar x_0, x_\Delta; |\overline{\Delta}|)$ for each
$\Delta$ close to $\Delta_\infty$, such that \be \label{liminf}
\lim_{\Delta \nearrow \Delta_\infty} V_\Delta - \J_{\bar
u_0}^{\overline{\Delta}}(y_\Delta) = \infty. \ee
 Concretely,  $y_\Delta$
will be taken as a minimizing geodesic. In fact, for some
constants $C_1, C_2 >0$ (and assuming
$\overline{x}_{0}=\overline{x}$ in (\ref{dist}) without loss of
generality)
$$
2\J_{\bar u_0}^{\overline{\Delta}}(y_\Delta) =
\frac{|x_\Delta|^2}{|\bar\Delta|} - \int_0^{|\overline{\Delta}|}
F(y_\Delta(s), \bar u_0 + \nu s ) ds \leq C_1 |x_\Delta|^2 + C_2 ,
$$
the equality by taking into account the minimizing character of
$y_{\Delta}$, and the inequality by the at most quadratic bound of
$|F|$ and the fact that $|\Delta|$ is bounded
($|\Delta_\infty|<\infty$). Therefore, the mentioned inequality $2
V_\Delta \geq  A_\Delta$ plus   Corollary \ref{wirtinger's2} (its
last assertion) yields
 the required limit (\ref{liminf}).

{\em (ii)}. As we have seen $V_\Delta$ diverges and, thus, it is
enough to prove the existence of $\nu \delta
>0$ such that $V_\Delta(\bar x_0, \bar u_0)$ is lower bounded for
any $\nu \Delta \in (0, \nu \delta)$. From the at most quadratic
behaviour of $F$ we have
 $$2\J_{\bar u_0}^{\bar\Delta}(y)
 \geq
\int_{0}^{|\overline{\Delta}|}|\dot{y}(s)|^2
ds -\int_{0}^{|\overline{\Delta}|}
(R_1 |y(s)|^2
+R_0) ds
$$
for any $y\in {\cal C}(\bar x_0, x_\Delta ; |\overline{\Delta}|)$
and for $\Delta$ such that $\nu \Delta \geq \nu (\Delta_\infty
-1)$ (so that the coefficients $R_1(u), R_0(u)$ for at most
quadraticity can be replaced by their maximums in $u$). Then, the
required $\delta$ and lower bound are straightforward from
Proposition \ref{pwirt}. \cvd

Notice that part (i) of Lemma \ref{previo} plus  Proposition
\ref{l-I-} yield:

 $$I^{-\nu}[\gamma]= \{ \overline{z}_{0}\in\M: \nu \overline{u}_{0}<
\nu u_{\infty} ( <\infty ) \},$$ and the part (ii) joined to
Proposition \ref{l-I-f} yield:
$$\uparrow^{\nu}\gamma \subset \{\bar z_0\in\M: \nu \bar u_0 > \nu (u_\infty +
\delta) \},
$$
which is an information additional to Theorem \ref{t-pastsets}.

Summarizing, the two ambient hypotheses of Lemma \ref{previo}
yield the following strengthening of the conclusions of Theorem
\ref{t-pastsets}.

\begin{theorem}\label{3} Let $\M$ be a Mp-wave with  $|F|$  at most quadratic
and $M$ complete. Choosing $(u_0, v_0)\in \R^2$,  $x: [0,
|\Delta_\infty|) \rightarrow M$  and $b^-$ as in Theorem
\ref{t-pastsets}, the equalities for non-empty past and future
sets read as:
\[
\begin{array}{lrr}
P=\left\{\begin{array}{rcl} \overline{z}_{0}: &
\overline{u}_{0}<u_{\infty} & {\rm if}\quad u_\infty <\infty
\\ \overline{z}_{0}: & b^-(\overline{x}_{0},\bar u_0)>\overline{v}_{0}-v_{0} & {\rm if} \quad u_\infty = \infty,
\end{array}\right.
&\quad &
\end{array}
\]
\[
\begin{array}{lrr}
F=\left\{\begin{array}{rcl} \overline{z}_{0}: &
\overline{u}_{0}>u_{\infty}
 & {\rm if}\quad u_\infty > -\infty \\
\overline{z}_{0}: & b^-(\overline{x}_{0},\bar u_0)> v_{0}-
\overline{v}_{0} & {\rm if} \quad u_\infty =-\infty.
\end{array}\right. 
& \quad &

\end{array}
\]
Even more, for each $P$, $F$ as above there exists $\nu\delta >0$
such that:
$$
\uparrow P= I^+[\{\overline{z}_{0}: \;  \overline{u}_{0}\geq
u_{\infty}+\nu\delta\; \; \hbox{and}\; \;
b^+(\overline{x}_{0},\bar u_0)\leq \overline{v}_{0}-v_{0}\}\subset
\{\bar z_0:  \bar u_0 > u_\infty + \nu\delta \},
$$
$$
(\hbox{resp.} \; \downarrow F= I^-[\{\overline{z}_{0}: \;
\overline{u}_{0} \leq u_{\infty}-\nu\delta\; \; \hbox{and}\; \;
b^+(\overline{x}_{0},\bar u_0)\leq v_{0}-\overline{v}_{0}\}\subset
\{\bar z_0:  \bar u_0 < u_\infty - \nu\delta \}),
$$
and, if $F$ is subquadratic, one can take $\delta = \infty$, i.e.:
$$
\uparrow P = \emptyset, \quad \quad \downarrow F= \emptyset.
$$
\end{theorem}

\begin{remark}\label{r1} {\em Notice that, in order to write the pairs $(P,F)\in \partial\M$,
a TIF $F$ cannot be $S$-related with two TIP's $P, P'$. In fact,
the corresponding $u_\infty$ should be finite for $P$ and $P'$
(otherwise, the common future would be empty) and, thus, one of
them, say $P$, would be included in the other, $P'$ (contradicting
the maximality of $P$ in $\downarrow F$).

In the subquadratic case for $|F|$, $\partial\M$ is the union of
all the pairs $(P,\emptyset)$ and $(\emptyset, F)$; in particular,
no ideal points in $\hat{\partial} \M$ and $\check{\partial} \M$
are identified (this is a general fact, for globally hyperbolic
spacetimes \cite[Th. 9.1]{F}). In the general at most quadratic
case, pairs $(P,F)$ with none of the components empty are allowed
(as well as identifications between points in $\hat{\partial} \M$
and $\check{\partial}\M$). But, even in this case, a non-empty $P$
can form an ideal point with at most one $F$, and viceversa.}
\end{remark}

\section{Mp-waves with natural 1-dim. $\partial \M$}\label{s8}

\subsection{Collapsing to $i^\pm$}\label{s8.1}

We begin by studying the case of lightlike curves with diverging
component $u$.

\begin{proposition}\label{propo!!} Let $\M$ be a Mp-wave with $|F|$ at
most quadratic, $M$ complete, and satisfying condition {\bf (H1)}
in Def. \ref{3.3}. If $\gamma:[0,\infty)\rightarrow \M$ is a
$\nu$-lightlike curve, 
then
\[
I^{-\nu}[\gamma]=\M,\quad \uparrow^\nu \gamma=\emptyset.
\]
\end{proposition}
{\it Proof.} Clearly, the second equality directly follows from
Theorem \ref{3}. For the fist one, fix $\overline{z}_{0}\in \M$.
Again from Theorem \ref{3}, it suffices to show that
$b^-(\overline{x}_{0},\overline{u}_{0})=\infty$. To this aim, we
only need to prove (recall (\ref{bemol-})): \be \label{kkk}
V_\Delta(\bar x_0, \bar u_0) = (\hbox{inf}_{{\cal C}}{\cal
J}_{\overline{u}_{0}}^{\overline{\Delta}}=) -\infty\qquad\hbox{for
all}\;\; \Delta \;\;\hbox{big enough} 
\ee with $\overline{\Delta}=\Delta +u_{0}-\overline{u}_{0}$,  and
${\cal C}\equiv {\cal C} (\bar x_0, x_\Delta;
|\overline{\Delta}|)$. Choose all the $\Delta$'s such that
$\overline{\Delta}-2$ is greater than the value of
$\Delta_{0}=\Delta_{0}(\overline{u}_{0}+1)$ given by hypothesis
{\bf (H1)}. Consider the following constant speed smooth curves:
$\alpha:[0,1]\rightarrow M$ joining $\overline{x}_{0}$ to
$\overline{x}$ and
$\beta_{\Delta}:[\overline{\Delta}-1,\overline{\Delta}]\rightarrow
M$ connecting $\overline{x}$ to $x(\Delta)$. Let $x_{m}$ be the
sequence of piecewise smooth loops provided by hypothesis {\bf
(H1)} for $u_0=\overline{u}_{0}+1$ and $\Delta u=
\overline{\Delta}-2$. The sequence of juxtaposed curves
$y_{\Delta_m} = \beta_\Delta \star x_m \star \alpha$, i.e.,
\[
y_{\Delta_m}(s)=\left\{\begin{array}{ll} \alpha (s) &
\hbox{if}\; s\in [0,1] \\
x_{m}(s-1) & \hbox{if}\; s\in
[1,\overline{\Delta}-1] \\
\beta_{\Delta}(s) & \hbox{if}\; s\in
[\overline{\Delta}-1,\overline{\Delta}],
\end{array}\right.
\]
satisfies:
\[
\begin{array}{rl}
{\cal J}_{\overline{u}_{0}}^{\overline{\Delta}}(y_{\Delta_m}) &
=\frac{1}{2}\int_{0}^{\overline{\Delta}} |\dot y_{\Delta_m}(s)|^2
ds -\frac{1}{2}\int_{0}^{\overline{\Delta}}
F(y_{\Delta_m}(s),\overline{u}_{0}+s)ds
\\ & =\frac{1}{2}{\rm length}(\alpha)^{2}+\frac{1}{2}{\rm
length}(\beta_{\Delta})^{2} \\ & \quad
-\frac{1}{2}\int_{0}^{1}F(\alpha(s),\overline{u}_{0}+s)ds -
\frac{1}{2}\int_{\overline{\Delta}-1}^{\overline{\Delta}}F(\beta_{\Delta}(s),\overline{u}_{0}+s)ds
\\ & \quad +{\cal
J}_{\overline{u}_{0}+1
}^{\overline{\Delta} - 2
}(x_{m}).
\end{array}
\]
Thus, hypothesis {\bf (H1)} ensures that ${\cal
J}_{\overline{u}_{0}}^{\overline{\Delta}}(y_{\Delta_m})$ goes to
$-\infty$ when $m$ goes to $+\infty$, and (\ref{kkk}) holds, as
required. \cvd

With this result and Theorem \ref{3} at hand, our aim in the next
subsections is to formalize precisely the cases when the boundary
of the wave is a lightlike line.

\subsection{Case asymptotically quadratic}\label{s8.2}

Now, if we take into account the boundary construction in
Subsection \ref{s3.2}, we can establish the following result:

\begin{theorem}\label{th-main} The causal boundary $\partial\M$ of a Mp-wave with $F$ $\lambda$-asymptotically quadratic for some $\lambda>1/2$, and $M$ complete has the following structure:

(a) As a point set, two copies $L^+, L^-$ of $\R$, with eventual
identifications between the points of the copies, plus two ideal
points $i^+$, $i^-$. In fact, $\partial\M$ will be written as a
union (non-necessarily disjoint, due to the identifications)
$\partial\M= \hat \partial\M \cup \check \partial\M$ where
 $ \hat \partial\M \equiv L^+ \cup \{i^+\}$ and $\check \partial\M \equiv \{i^-\} \cup L^-$.

(b) Topologically, the following natural homeomorphisms hold:
$\hat
\partial\M  \cong   (-\infty, \infty]$, $\check
\partial\M \cong  [-\infty, \infty)$. Moreover, $\partial\M$
is a quotient topological space with the possible identifications
allowed in (a) above.

(c) Causally, $\hat{\partial}\M$, $\check \partial\M$, with the
restriction of the weak causal relation in $\partial\M$, are
totally ordered and weakly locally lightlike (i.e., each $Q$ in,
say, $\hat
\partial\M$ has a neighbourhood ${\cal L} \subseteq  \hat \partial\M$   such that: any $Q_1, Q_2 \in {\cal L}$ are weakly horismotically
related in $\hat
\partial\M$ if and only if  $Q_1 < Q_2$ as points of $(-\infty,
\infty]$).
\end{theorem}
{\it Proof.} From Lemma \ref{l22} {\it (i)} and Definition
\ref{superqu} {\it (iii)},  the hypotheses of Theorem \ref{3},
Proposition \ref{propo!!} hold. Therefore, directly from
Proposition \ref{propo!!} and Theorem \ref{3}:
\[
I^{-}[\gamma]=\M,\qquad
\uparrow\gamma=\emptyset\qquad\qquad\qquad\qquad\qquad\qquad\qquad\hbox{if}\;\;
\Delta_{\infty}=\infty,
\]
\be \label{ebemoles} \left\{\begin{array}{l}
I^{-}[\gamma]=\{\overline{z}_{0}:\;\;\overline{u}_{0}<u_{\infty}\} \\
\uparrow\gamma=I^{+}[\{\overline{z}_{0}:\; \overline{u}_{0}\geq
u_{\infty}+\delta,\;\;
b^+(\overline{x}_{0},\overline{u}_{0})+v_{0}-\overline{v}_{0}\leq
0\}]
\end{array}\right. \qquad \hbox{if}\;\; \Delta_{\infty}<\infty,
\ee for any future-directed lightlike curve $\gamma$ with
$u(s)=u_{0}+s$. Thus, the future causal boundary $\hat{\partial}
\M$ contains the ideal point $i^{+}$ and a copy $L^{+}$
corresponding to the line $u_{\infty}\in (-\infty,\infty)$.
Moreover, the chronological topology clearly attaches $i^{+}$ to
the right extreme of $L^{+}$ (and it is the natural topology on
$L^+$). On the other hand, any two points $u_{\infty},
u'_{\infty}\in L^{+}$, $u_{\infty}<u'_{\infty}$, are weakly
causally related, since the corresponding pairs of terminal sets
$(P,F)$, $(P',F')$ satisfy:
\begin{equation} \label{fweak}
P=\{\overline{z}_{0}: \overline{u}_{0}<u_{\infty}\}\subset
\{\overline{z}_{0}:\overline{u}_{0}<u'_{\infty}\}=P'.
\end{equation}
Moreover, taking into account that $F\subset \uparrow P\subset
\{\overline{z}_{0}: \overline{u}_{0}> u_{\infty}+\delta\}$ for
some $\delta>0$ (recall Theorem \ref{3}), one has, for
$u_{\infty}<u'_{\infty}\leq u_{\infty}+\delta$,
\[
F\cap P'\subset\{\overline{z}_{0}: u_{\infty}+\delta
<\overline{u}_{0}<u'_{\infty}\}=\emptyset.
\]
Whence, $(P,F)$, $(P',F')$ are not chronologically related, and
$\hat
\partial\M$ is weakly locally lightlike.

Analogously, the past causal boundary $\check{\partial}\M$ can be
represented by another copy $L^{-}$ of the line $u_{\infty}\in
(-\infty,\infty)$ plus the ideal point $i^{-}$ attached at the
left extreme, and is weakly locally lightlike.

Finally, the (total) causal boundary $\partial\M$ is formed by
$L^{+}\cup \{i^{+}\}\cup L^{-}\cup \{i^{-}\}$, up to eventual
identifications between those ideal points in $L^{-}$, $L^{+}$
represented by the same pair of terminal sets, and all the
conclusions follow. \cvd

\begin{remark} \label{rweak} {\em Notice that we have stated only the weak causal
relation, as  we have proven $P\subset P'$ in (\ref{fweak}) but
not $F'\subset F$. The possible difficulty for this inclusion
appears only in the very particular case that $F'$ is a maximal
TIF into $\uparrow P$, and $P'$ a maximal TIP into $\downarrow
F'$, and thus $F=\emptyset$. This situation cannot happen if, for
example, $F(x,u)$ is independent of $u$, since then $F$ is maximal
TIF into $\uparrow P$ if and only if $P$ is maximal TIP into
$\downarrow F$. As a consequence, the boundary in this case
becomes locally lightlike for the natural causal
relation.}\end{remark}

\subsection{Plane waves}\label{s8.3}

Consider now the case of a plane wave $\M=\R^{n}\times \R^{2}$,
\[
F(x,u)= \sum _{i,j}f_{ij}(u)x^{i}x^{j}, \quad \quad f_{ij}=f_{ji}.
\]
For simplicity, assume that $F$ has the form of Lemma \ref{l22}
{\it (ii)} and, thus, falls under the hypotheses of Th. \ref{3}
and Prop. \ref{propo!!}. Then, reasoning as in Th. \ref{th-main}:
\begin{theorem}\label{previous}
The causal boundary of a plane wave with $ f_{1j}\equiv 0$ for all
$ j\neq 1,$ and $f_{11}(u)\geq\lambda^{2}/(u^{2}+1)$, for large
$|u|$ and some $\lambda>1/2$, is as described in Th.
\ref{th-main}, Remark \ref{rweak}.
\end{theorem}

\begin{remark} \label{rcagna} {\em
Some particular cases where $f_{ij}$ is diagonal have been
computed by Hubeny and Rangamani in \cite{HR}, and it is worth
comparing here. They used the existence of ``oscillating
geodesics'' as an evidence of a 1-dimensional boundary. The items
in \cite[Subsection 4.3]{HR} labelled 1, 2,  NL1, NL3 as well as
the case $f_{11}(u)=1/(u^{2}+1)$ of item 4 (or the singular case
NL2) do have such oscillating geodesics, and are particular cases
of Th. \ref{previous}. The case $f_{11}(u)=\cos u$ (item 3), is
included in the technique, as it satisfies trivially the
inequality (\ref{sturm}) and, thus the conclusion of Lemma
\ref{l22} holds (see Remark \ref{rsturm}). In the singular case
$f(u)=\lambda^2/u^2$ (item 6) they obtain oscillatory geodesics
for $\lambda^2>1/4$, also in agreement with Th. \ref{previous}. As
shown in Subsection \ref{s9.1} by means of a counterexample, one
cannot expect a 1-dimensional boundary even in the limit case
$\lambda^2=1/4$. So, it is not surprising now that, if
$f_{11}(u)=e^{-u^2}$ (as in \cite[Subsect. 4.3, item 5]{HR}) the
oscillatory behaviour ceases.

Very roughly, in our approach  the infimum of some functional is
considered, and in Hubeny and Rangamani's just the (lightlike
geodesics associated to the) critical curves of this functional.
Of course, when the infimum is attained the minimizing curve is
critical, but our functional approach has clear advantages. In
fact, it relies only on the qualitative functional properties
rather than on the exact details of the Euler-Lagrange equation.
The oscillating geodesics in the most accurate  Hubeny and
Rangamani's results, imply the existence of a solution with two
zeros for the Euler-Lagrange equation of our simplified functional
(\ref{simple}) (see the discussion around this formula), and this
is enough for the results.}
\end{remark}

Recall that only the 1-dimensional character of the boundary is
ensured by  Th. \ref{previous}, \ref{th-main}. The question of
establishing which ideal points in $L^{+}$ and $L^{-}$ must be
identified becomes hard, and depends on the behaviour of function
$b^+$ in (\ref{ebemoles}). The only additional information on
$b^+$ is provided by Lemma \ref{previo}(ii) (or, equivalently, by
the expressions of $\uparrow P, \downarrow F$ in Th. \ref{3}).

Nevertheless, identifications can be easily computed in the highly
symmetric case of plane waves with $F(x,u)$ independent
of\footnote{They are usually called {\em homogeneous} plane waves,
even though the name {\em locally symmetric} is intrinsic and
seems more appropriate, see for example \cite{GP-Se}.} $u$, i.e.,
$F(x,u)=\sum_{ij}\mu_{ij}x^{i}x^{j}$, with $\mu_{ij}$ symmetric
coefficient matrix. Here, each $\uparrow P$ is equal to some $F$
and viceversa \cite{MR}. As remarked in \cite{MR}, these Mp-waves
contain many interesting examples for string theory (maximally
supersymmetric 11-dimensional solution obtained from the Penrose
limit of $AdS_{4}\times S^{7}$ and $AdS_{7}\times S^{4}$
\cite{BFHP}, partially supersymmetric plane waves in ten
dimensions \cite{CLP, BR, Mi}, including the Penrose limit of the
Pilch-Warner flow \cite{CHKW, GPS, BJLM}). Due to the
exceptionality of this case, we will not attempt a very general
result here. Simply, we will give an extended version of the
result in \cite{MR}, in order to check how our technique works.
More general results would rely on the possibility to reformulate
Lemma \ref{5.4} below and extend formulas (\ref{kkkk}),
(\ref{kkkkk}).

Concretely, now we assume that function $f_{11}$ in Th.
\ref{previous} is constant and equal to the biggest eigenvalue
$\mu_1$ of the matrix $f_{ij}(u)$, and $\mu_1>0$.

\begin{lemma}\label{5.4} Under these hypotheses, let $\gamma:[0,|\Delta_\infty|)\rightarrow
\M$ be an inextendible $\nu$-lightlike curve, with $|\Delta_\infty
|\in (0,\infty)$, $\nu \Delta_\infty >0$ and
$u_\infty:=u_{0}+\Delta_\infty$. Then:

If $\nu=1$, $\uparrow \gamma=\R^{n}\times
(u_\infty+\pi/\mu_{1},\infty)\times \R$.

If $\nu=-1$,  $\downarrow \gamma=\R^{n}\times
(-\infty,u_\infty-\pi/\mu_{1})\times \R$.
\end{lemma}
{\it Proof.} (For $\nu=1$.) $\supseteq$. Clearly, if
$\overline{z}'_{0}\in \R^{n}\times
(u_\infty+\pi/\mu_{1},\infty)\times \R$ then $\overline{z}'_{0}\gg
\overline{z}_{0}$ for some
$\overline{z}_{0}=(\overline{x}_{0},\overline{u}_{0},\overline{v}_{0})$
with $\overline{u}_{0}=u_\infty+\pi/\mu_{1}$. Therefore, from
Proposition \ref{l-I-f} the required inclusion  follows by proving
$b^+(\overline{x}_{0}, \bar u_0=u_{\infty}+\pi/\mu_1)=-\infty$, or
just (recall (\ref{bemol+})):

\be \label{kkkk} V_\Delta(\bar x_0, \bar u_0)=-\infty \qquad
\hbox{for all} \; \Delta < \Delta_\infty \; \hbox{close to} \;
\Delta_\infty.\ee Thus, for $\Delta$ close to $\Delta_\infty$,
consider $|\overline{\Delta}|(>\pi /\mu_1)$ as in (\ref{incr}) and
take $0<\delta_\Delta <|\overline{\Delta}|/2$ small enough such
that
\begin{equation}\label{cond-ineq}
\mu_{1}^{2}
\geq\frac{\pi^{2}+\epsilon_\Delta}{(|\overline{\Delta}|-2\delta_\Delta)^2},\qquad\hbox{for
some}\;\;\epsilon_\Delta>0.
\end{equation}
Define the juxtapositions
\[
y_{\Delta_m}(s)=\left\{\begin{array}{ll} -\frac{x(\Delta)}{\delta_\Delta}s+x(\Delta) & \hbox{if}\; s\in [0,\delta_\Delta] \\
(y^1_{\Delta_m}(s),0,\ldots,0) & \hbox{if}\; s\in
[\delta_\Delta,|\overline{\Delta}|-\delta_\Delta]
\\ \frac{\overline{x}_{0}}{\delta_\Delta}s+\frac{\delta_\Delta \overline{x}_{0}-|\overline{\Delta}|\overline{x}_{0}}{\delta_\Delta} & \hbox{if}\; s\in
[|\overline{\Delta}|-\delta_\Delta,|\overline{\Delta}|],
\end{array}\right.
\]
with
\[
y^1_{\Delta_m}(s)=
m\sin\left(\frac{\pi}{|\overline{\Delta}|-2\delta_\Delta}(s-\delta_\Delta)\right)
\quad \quad \forall s\in
[\delta_\Delta,|\overline{\Delta}|-\delta_\Delta].
\]
Then, from (\ref{cond-ineq}) we obtain
\[
\begin{array}{rl}
{\cal J}_{u_\Delta}^{\overline{\Delta}}(y_{\Delta_m}) &
=\frac{1}{2}\int_{0}^{|\overline{\Delta}|}(|\dot
y_{\Delta_m}(s)|^2 -F(y_{\Delta_m}(s),u_\Delta +s))ds \\ & =
\frac{1}{2}\left(\int_{\delta_\Delta}^{|\overline{\Delta}|
-\delta_\Delta}|\dot y_{\Delta_m}^{1}(s)|^{2} ds
-\mu_{1}^{2}\int_{\delta_\Delta}^{|\overline{\Delta}|-\delta_\Delta}y^1_{\Delta_m}(s)^{2}ds\right)+\Lambda_\Delta
 \\ & \leq
\frac{1}{2}\int_{\delta_\Delta}^{|\overline{\Delta}|-\delta_\Delta}|\dot
y_{\Delta_m}^{1}(s)|^{2} ds
-\frac{\pi^{2}+\epsilon_\Delta}{2(|\overline{\Delta}|-2\delta_\Delta)^{2}}\int_{\delta_\Delta}^{|\overline{\Delta}|-\delta_\Delta}y^1_{\Delta_m}(s)^{2}ds+\Lambda_\Delta
 \\ & =-\frac{\epsilon_\Delta m^{2}}{4(|\overline{\Delta}|-2\delta_\Delta)}+\Lambda_\Delta
\end{array}
\]
for some $\Lambda_\Delta\in\R$ independent of $m$. Summing up,
${\cal J}_{u_\Delta}^{\overline{\Delta}}(y_{\Delta_m})\rightarrow
-\infty$ when $m \rightarrow +\infty$, and (\ref{kkkk}) holds.

 $\subseteq$. We will prove that, if $\overline{z}'_{0}\not\in
\R^{n}\times (u_\infty+\pi/\mu_{1},\infty)\times \R$ then
$\overline{z}_{0}\not\in \cap_{\Delta}I^{+}[\gamma(\Delta)]$ for
any $\overline{z}_{0}\ll \overline{z}'_{0}$ (and thus,
$\overline{z}'_{0}\not\in \uparrow\gamma$). From Lemma \ref{l1},
$\overline{u}_{0}-u_\infty<\pi/\mu_{1}$, and by Prop. \ref{l-I-f},
it is enough:
 \be \label{kkkkk} V_\Delta(\bar
x_0, \bar u_0)
>-\infty\quad\hbox{is
lower bounded for all} \; \Delta <\Delta_\infty \; \hbox{close
to}\; \Delta_\infty \ee (recall (\ref{bemol+}) and the fact that
$V_\Delta \rightarrow \infty$ because
 of  (\ref{des-def}) and Cor. \ref{wirtinger's2}). From
the hypotheses, $|\overline{\Delta}| \leq
(\pi-\epsilon_{0})/\mu_{1}$, for some $\epsilon_{0}>0$, and for
all $\Delta <\Delta_\infty$ close enough. Therefore,
\[
\begin{array}{rl}
{\cal J}_{u_{\Delta}}^{\overline{\Delta}}(y) &
=\frac{1}{2}\int_{0}^{|\overline{\Delta}|}(|\dot{y}(s)|^2
-F(y(s),u_{\Delta}+s))ds \\ & \geq
\frac{1}{2}\left(\int_{0}^{|\overline{\Delta}|}|\dot{y}(s)|^2
ds-\mu_{1}^{2}\int_{0}^{|\overline{\Delta}|}|y(s)|^{2}ds\right)
\\ & \geq\frac{1}{2|\overline{\Delta}|}\left(|\overline{\Delta}|\int_{0}^{|\overline{\Delta}|}|\dot{y}(s)|^2
ds-\frac{(\pi-\epsilon_{0})^{2}}{|\overline{\Delta}|}\int_{0}^{|\overline{\Delta}|}|y(s)|^{2}ds\right).
\end{array}
\]
As $V_\Delta(\bar x_0, \bar u_0)$ is obtained by taking the
infimum in this expression, the bound for $\lambda$ in Theorem
\ref{wirtinger's} (see Appendix) ensures (\ref{kkkkk}), as
required. \cvd

\vspace{2mm}

Theorem \ref{3} and Lemma \ref{5.4} tell us that the pair
$(I^-[\gamma], \uparrow \gamma)$ with $u \nearrow u_\infty$
coincides with $(\downarrow \tilde \gamma, I^+[\tilde \gamma]))$
with $u \searrow u_\infty+\pi/\mu_{1}$, i.e., each future ideal
point represented by some $u_{\infty}\in L^{+}$ must be identified
with the past ideal point represented by
$u_{\infty}+\pi/\mu_{1}\in L^{-}$ (and there are no more
identifications). Summing up:
\begin{theorem}
Let $\M$ be a plane wave with $f_{1j}\equiv 0$ for all $ j\neq 1,$
and $f_{11}(u)$ a positive constant function equal to the biggest
eigenvalue of $f_{ij}(u)$ (in particular, any locally symmetric
plane wave with a positive eigenvalue). Then, $\partial \M$ is
weakly locally lightlike and canonically identifiable to
$[-\infty, \infty]$, both as a point set and as a topological
space, being the weak causal relation the corresponding one to the
natural order. Even more, in the locally symmetric case this also
holds for the causal relation.
\end{theorem}

\section{Higher dimensionality of $\partial \M$}\label{s9}

When $F$ grows less fast than quadratic (in all directions) one
does not expect a 1-dimensional boundary. In fact, if $F$ is
subquadratic and $M$ complete then the Mp-wave becomes globally
hyperbolic. So, there are no identifications between
$\hat{\partial}\M$, $\check{\partial}\M$ and, the structure of the
spacetime suggests a boundary with two pieces which resemble in
some sense the Cauchy hypersurfaces\footnote{If $M$ were not
complete, global hyperbolicity may be destroyed, but the main
difference in the expected picture is that additional boundary
points would appear, associated to inextendible curves in $M$ with
finite energy.} --notice that the Cauchy hypersurfaces are
necessarily noncompact and, at least when $M$ is non-compact, one
could expect that some portion of $\partial M$ were higher
dimensional, even of dimension $(n+1)$. Some concrete cases will
be briefly analyzed in Subsections \ref{s9.2}, \ref{s9.3}. But,
first, we will see that the ($\lambda=1/2$)-asymptotic quadratic
growth of $F$ becomes critical for the 1-dimensional character of
the boundary. Recall that this case appears in geometries derived
from NS5 branes, see \cite[Sect. 4.3, $\S$NL2]{HR}.

\subsection{Criticality of $\lambda =1/2$ for 1-dimensionality }\label{s9.1}

Consider for simplicity a pp-wave $\M = \R^{n+2}$ with
$F=F_\lambda, \lambda\in \R$, satisfying: \be \label{estar}
F_\lambda(x,u)=\lambda^{2}|x|^{2}/(1+u)^{2} ,\ee for $ u\geq 0$
(and eventually for $u<-2$, but we will not take care of this
part). Obviously, $F_\lambda$ is $\lambda$-asymptotically
quadratic and, for  $\lambda>1/2$, $\hat{\partial}\M$  is
1-dimensional (and so essentially $\partial \M$). Our purpose is
to show that this is {\em not} the case for $\lambda=1/2$, which
shows the optimal character of our results.

Concretely, we will construct $\nu$-lightlike curves $\gamma :
[0,\infty) \rightarrow \M$  with $u(s)\nearrow \infty$ such that
$I^{-}[\gamma]\neq \M$. Thus, the collapse of all the
corresponding ideal points to the single one $i^+$ (which was
essential in Section \ref{s8} --Prop. \ref{propo!!}-- in order to
ensure the  1-dimensionality of the boundary) will not hold. As a
technical previous step:

\begin{lemma} \label{mascagna}
Let $F=F_{1/2}$ in (\ref{estar}) and $n=1$. Consider the
functional
$${\cal J}_0^{\Delta u}(x)=\int_0^{\Delta u}\left( \dot x^2-F(x(u),u)\right)
du$$ and the solution $y(u)= \sqrt{1+u}$ to the Euler-Lagrange
equation
$$
\ddot y + p(u) y = 0 \quad \quad p(u)=1/4(1+u)^2.
$$
Then
$$
\mbox{{\rm Inf}}_{x\in {\cal C}(1,y(\Delta u);\Delta u)}{\cal
J}_0^{\Delta u} = {\cal J}_0^{\Delta u}(y|_{[0,\Delta u]})=0$$ for
all $\Delta u >0$.
\end{lemma}

{\em Proof.} The last equality is straightforward, so, we will see
that $y|_{[0,\Delta u]}$ minimizes the functional by usual
techniques from Sturm-Liouville theory (see \cite[Sect.
1.1]{Beesack}, \cite[Ch. 4]{Ze}). Put $g(u)=\dot y(u)/y(u)$, which
satisfies Riccati's equation $\dot g + g^2=-p$. For any $x\in
{\cal C}(1,y(\Delta u);\Delta u)$ one has: \be \label{eeeee} {\cal
J}_0^{\Delta u}(x)=\int_0^{\Delta u}\left( \dot x^2-   p
x^2\right) du = \int_0^{\Delta u}\left( \dot x-  x g\right)^2 du +
\left. x^2(u) g(u)\right]^{u=\Delta u}_{u=0}\ee (expand the first
term in the right side and integrate by parts $2\int  x \dot x g =
\int \dot{x^2} g$). And taking into account that curves $x, y$
coincide at the extremes:
$${\cal J}_0^{\Delta u}(x)\geq  \left. x^2(u)
g(u)\right]^{u=\Delta u}_{u=0} =  \left. y^2(u)
g(u)\right]^{u=\Delta u}_{u=0} = {\cal J}_0^{\Delta u}(y)$$ (the
last equality applying (\ref{eeeee}) to $x=y$).\cvd

Now, consider the lightlike curve in the pp-wave
$\gamma(u)=(x(u),u,v(u))$ constructed from Lemma \ref{primaria}
with $x(u)= y(u) \vec e$, where $\vec e$ is any unit vector of
$\R^n$ and  $y(u)= \sqrt{1+u}$ (and $v(0)=0$). From (\ref{V}) and
Lemma \ref{mascagna}, the arrival function $V$ satisfies:
$$
V((x(0),0), (x(u), u))= 0, \quad \forall u>0.
$$
Thus, from the interpretation of $V$ (Prop. \ref{l2}),  $z=(x(0),
0, v_0) \not\in I^-(\gamma(u))$ whenever $v_0\geq 0(=v(u))$.

\begin{remark} {\em Notice that this not only proves the required inequality
$I^{-}[\gamma]\neq \M$. In fact, moving $\vec e$ in all the
directions ($\vec e \in \SSS^{n-1}\subset \R^n$), and $v(0)\in
\R$, different curves $\gamma= \gamma[\vec e, v(0)]$ are obtained.
Each one yields an ideal point, that is, a portion of $\partial
\M$ containing a $n$-dimensional subset of ideal points is
constructed.}
\end{remark}

\subsection{Static and Minkowski type Mp-waves} \label{s9.2}

According to Garc\'{\i}a-Parrado and Senovilla \cite{GP-Se1}, a
spacetime $\M$ is called {\em causally related} with a second one
$\M '$, shortly $\M \prec \M '$, if a diffeomorphism $\phi$ maps
the causal cones of $\M$ into the ones of $\M'$; moreover, $\M, \M
'$ are {\em isocausal} if $\M \prec \M '$ and $\M' \prec \M$.
Intuitively, when $\M \prec \M '$ the causal  cones of $\M '$ can
be obtained by opening the ones of $\M$. If they are isocausal
then they are not necessarily conformal, but many causal
properties are shared by both spacetimes \cite{GP-Se1, GP-Sa}.

When a Mp-wave has coefficient $F(x,u)$ bounded in $x$ then it
becomes isocausal to the simplest choice $F\equiv 0$, more
precisely:
\begin{proposition} \label{isocausal}
Let $(\M, \langle \cdot , \cdot \rangle_L)$ be a Mp-wave with
$|F(x,u)|\leq f(u)$ for all $(x,u)\in M\times \R$, where  $f$ is a
continuous function. Then $(\M, \langle \cdot , \cdot \rangle_L)$
is {\em isocausal} to the standard static spacetime obtained just
making $F\equiv 0$, i.e.
$$
 \M = M\times \R^2, \quad g_0= \langle \cdot , \cdot \rangle -2dudv.
$$
\end{proposition}
{\em Proof.} By a simple computation of the causal cones, the
metrics
\[
g^{\pm}:=\langle\cdot,\cdot\rangle \pm f(u)du^{2}-2dudv
\]
satisfy
$$ (\M, g^-) \prec (\M, \langle \cdot , \cdot \rangle_L) \prec
(\M,g^+).$$ But recall that both metrics $g^\pm$ are isometric to
the static $(\M, g_0)$, as shown by the global change of
coordinates:
\[
\tilde u = u, \quad \quad
\tilde{v}=v\mp\frac{1}{2}\int_{0}^{u}f(\sigma)d\sigma.
\]
\cvd

 So, even though the relation between the causal
boundaries of two isocausal spacetimes does not seem trivial, one
expects that, when Proposition \ref{isocausal} applies, the
boundary of the Mp-wave will not be too different to the boundary
of the corresponding static model. In particular,  when $M = \R^2$
the static spacetime is $\LL^{n+2}$, so, if pp-waves are
considered, one expects a  boundary not very different to
Lorentz-Minkowski's.

\subsection{The case $-F$ quadratic} \label{s9.3}

Marolf and Ross \cite{MR} proved that the {\em conformal} boundary
is a set of two lightlike hyperplanes joined by two lightlike
lines, in the case of (conformally flat) locally symmetric plane
waves with equal negative eigenvalues. Now,  we will extend that
proof to include non-locally symmetric ones. Then, the causal
boundary will be also computed and, as we will see, the picture
will be a bit different.

We will also focus on the simplest case of a (non-locally
symmetric) plane wave with equal negative eigenvalues of $F$. This
corresponds to the case $-F$ quadratic (which can be studied in
further detail with the introduced techniques). Thus, let
$\M=\R^{n+2}$ with

\begin{equation}\label{simplified}
\begin{array}{l}
\langle\cdot,\cdot\rangle_{L}= dx^2 + |x|^2 f(u)du^{2} -2du\,dv
,\qquad f(u)>0,
\end{array}
\end{equation}
where, $x=(x^1,\dots , x^n)$. Consider the differential equation
\be \label{eee2} \ddot{r}(u)=f(u)r(u), \quad r(0)= 1, \quad
\dot{r}(0)=0. \ee The change of variables
\[
x= r(u)  \tilde{x}, \quad \quad v=
\tilde{v}+\frac{1}{2}r(u)\dot{r}(u)\tilde{x}^{2}
\]
takes (\ref{simplified}) into
\[
\langle\cdot,\cdot\rangle_{L} = r(u)^{2}d\tilde{x}^{2} -2du
d\tilde{v},
\]
on all $\R^{n+2}$. Thus, the further change of variable $
\tilde{u}=\int_{0}^{u}\frac{du'}{r(u')^{2}} $ yields the
explicitly conformally flat expression: \be
\label{min}\langle\cdot,\cdot\rangle_{L}
=r(\tilde{u})^{2}(d\tilde{x}^{2}-2d\tilde{u}d\tilde{v}). \ee
Observe that the domain for coordinate $\tilde{u}$ is given by:
\[
\tilde{u}_{-\infty}<\tilde{u}<
\tilde{u}_{\infty}\quad\hbox{with}\quad \tilde{u}_{\pm \infty}:=
\int_{0}^{\pm \infty}\frac{ds}{r(s)^{2}}, \quad 0< \pm \tilde
u_{\pm \infty} <\infty,
\]
being the finiteness of $\tilde{u}_{\pm \infty}$ a consequence of
the convexity of $r$ in (\ref{eee2}). Therefore, the plane wave is
conformal to the proper region
$\tilde{u}_{-\infty}<\tilde{u}<\tilde{u}_{\infty}$ of Minkowski
spacetime (in the coordinates of (\ref{min})).
 In particular, the conformal  boundary (for the restriction of the
 classical Minkowski embedding)
consists of two parallel lightlike hyperplanes at $\tilde{u}=\pm
\tilde{u}_{\infty}$ and two lightlike lines (say, two copies of
$[\tilde{u}_{-\infty}, \tilde{u}_\infty]$) which represent the
intersection of the region $\tilde{u}_{-\infty}\leq \tilde{u} \leq
\tilde{u}_{\infty}$ with the past and future infinity ${\cal
J}^\pm$ of Minkowski space.

Now, recall that the conformal version (\ref{min}) of the plane
wave (\ref{simplified}) can be also used to compute the causal
boundary, and it looks like  somewhat different. In fact, this
boundary contains again two lightlike hyperplanes (which can be
identified in $\LL^{n+2}$ with pairs $(I^-(z), \emptyset)\in
\hat{\partial}\M$, where $u(z)=\tilde{u}_\infty$, and $(\emptyset,
I^+(z))\in \check{\partial}\M$ with $u(z)=\tilde{u}_{-\infty}$),
and two lightlikes lines. But these lines are now identified
naturally with copies $(\tilde{u}_{-\infty}, \tilde{u}_\infty]
\subset \hat{\partial}\M$ and $[\tilde{u}_{-\infty},
\tilde{u}_\infty)\subset \check{\partial}\M$ (say, as no
future-directed timelike curve approaches $\tilde{u}_{-\infty}$).
Notice that, {\em both $\hat{\partial}\M$ and $\check{\partial}\M$
are connected and non-compact, and there are no identifications
for $\partial \M$; thus, plainly  $\partial
 \M = \hat{\partial}\M \cup
\check{\partial}\M$}.

\section{Conclusions}\label{s10}

We have carried out a systematic study of Mp-waves, being our main
goals:

\ben \item We consider the very wide family of wave-type
spacetimes (\ref{m-general}) and determine the general qualitative
behaviour of the metric which yields a 1-dimensional causal
boundary, as well as other properties, see Table 1.

\item Even though we particularize our general results to many
cases, our main aim is to introduce general techniques potentially
applicable to other cases of interest in  General Relativity,
String Theory or other theories. These techniques involve a
functional approach, Sturm-Liouville theory, the introduction of
new Busemann type functions and technicalities on Causality.

\item The functional approach (which is a variant of the one
introduced in \cite{FS}) is also interpreted as an arrival time
function, with clear analogues to Fermat's principle one. This
interpretation also clarifies the causal structure of the waves,
including the inexistence of horizons.

\item Our study includes the improvements on the notion of  causal
boundary in \cite{MR2, F}. Even though the well-known historical
problems of this notion can be minimized in a first approach (as
in \cite{HR}), finally a consistent notion of the identifications
of future and past sets, as well as a reasonable topology, must be
carried out. In fact, the former may lead to new interpretations
(in order to go {\em beyond infinity}, as claimed in \cite{MR})
and the latter is unavoidable to speak on the {\em dimension} of
the boundary. What is more, the new Busemann-type functions
$b^\pm$ here introduced seem to have general applicability for
this notion of causal boundary.

\een

Summing up, this work has  obvious contents for classical
Causality and General Relativity, and it is also introduced as a
tool for the string community, in order to check the exact
possibilities of holography on plane waves backgrounds.

\np

\begin{center}
\noindent
\begin{tabular}{|c|c|c|c|} \hline
 & & & \\
 \mbox{Qualitative $F$} &  \mbox{Causality} & \mbox{Boundary $\partial M$} & \mbox{Some examples}
  \\
 & & & \\ \hline
 & & & \\
$ \begin{array}{c} \mbox{$F$ superquad.} \\ \mbox{$-F$ at most
quad.}\end{array} $ & $\begin{array}{c} \mbox{No distin-} \\
\mbox{guishing}\end{array}$
 & \mbox{No boundary} & $\begin{array}{c} \mbox{pp-waves yielding} \\
\mbox{ Sine-Gordon string} \\ \mbox{and related ones}\end{array}$  \\
 & & & \\ \hline
 & & & \\
 $\begin{array}{c} \mbox{At most quad. $F$} \\
\mbox{(resp.$^1$ $|F|$)}\end{array}$
 & $\begin{array}{c} \mbox{Strongly} \\
\mbox{causal}\end{array}$ &  $\begin{array}{c} \mbox{Computable} \\
\mbox{from Th. \ref{t-pastsets}} \\
\mbox{(resp. Th. \ref{3})}
\end{array}$ & all below
  \\
 & & & \\ \hline
 & & & \\
$\begin{array}{c} \mbox{$\lambda$-asymp. quad.$^2$} \\
\mbox{$\lambda>1/2$}\end{array}$
 & $\begin{array}{c} \mbox{Strongly} \\
\mbox{causal}\end{array}$ & $\begin{array}{c} \mbox{1-dimension,} \\
\mbox{lightlike}\end{array}$ & $\begin{array}{c} \mbox{plane waves} \\
\mbox{with some eigenv.} \\
\mbox{$\mu_1 \geq \lambda^{2}/(1+u^2)$} \\ \mbox{for $|u|$ large }\end{array}$  \\
 & & & \\ \hline
 & & & \\
$\begin{array}{c} \mbox{$\lambda$-asymp. quad.} \\
\mbox{$\lambda \leq 1/2$}\end{array}$
 & $\begin{array}{c} \mbox{Strongly}
\\
\mbox{causal}
\end{array}$ &
$\begin{array}{c} \mbox{Critical}
\end{array}$ &
$\begin{array}{c} \mbox{pp-wave with}
\\
\mbox{$F(x,u)=\lambda^{2} x^2/(1+u)^2$}\\
\mbox{(for $ u>0$)}
\end{array}$
\\
 & & & \\ \hline
 & & & \\
 Subquadratic & $\begin{array}{c} \mbox{Globally} \\
\mbox{hyperbolic}\end{array}$ & $\begin{array}{c} \mbox{No identif.} \\ \mbox{in $\hat{\partial} \M, \check{\partial} \M$} \\ \mbox{Expected} \\
\mbox{higher dim.}\end{array}$ &
$\begin{array}{c} \mbox{ (1) $\LL^n$ and static} \\
\mbox{type Mp-waves}
\\
 \mbox{(2)  plane waves with} \\
\mbox{$-F$ quadratic}
\end{array}$  \\
 & & & \\ \hline
\end{tabular} \\
\vspace{\parskip}

\end{center}

\noindent {\small $^1$For this subcase and the cases below, assume
$M$  complete.}

\noindent  {\small $^2$It is sufficient for this asymptotic
behaviour to hold in a spatial direction of $M$ if $|F|$ is at
most quadratic. For other generalizations, see formula
(\ref{sturm}) and Remark \ref{rsturm}.}
\\

\begin{quote}
\begin{center}
{\sc Table 1} \\
\end{center}
Rough properties of the causal boundary of a Mp-wave depending on
the qualitative behaviour of $F$.
\end{quote}

\np

\appendix

\section{Appendix}

\begin{theorem}\label{wirtinger's} Let $M$ be a Riemannian
manifold and $x_m:[0,\Delta_m]\rightarrow M$ a sequence of
piecewise smooth curves with diverging energies
 and such that the endpoints $x_m(0), x_m(\Delta_m)$ are
 contained in a
bounded region $B$ of $M$ for all $m$. Then, for any
$\lambda<\pi^{2}$, and any $\mu,k\in\R$, $0<\epsilon<2$:
\[
\Delta_m\int_{0}^{\Delta_m}|\dot{x}_m(s)|^2 ds
-\frac{1}{\Delta_m}\int_{0}^{\Delta_m}(\lambda |x_m(s)|^2+\mu
|x_m(s)|^{2-\epsilon}+k)ds\rightarrow\infty.
\]
Moreover, if the assumption on the endpoints is done only for the
initial ones (i.e., $\{x_m(\Delta_m) \}_m$ does not lie
necessarily  in a bounded $B$) then the same assertion holds for
$\lambda<\pi^{2}/4$.
\end{theorem}
{\it Proof.} For each $m$, take the variable $\bar s = s/
\Delta_m$, $\bar x_m(\bar s) = x_m(\Delta_m \bar s)$ and write the
corresponding expression (up to a factor 2) as a typical
Lagrangian type kinetic minus potential energy:
\[
\frac{1}{2} \int_{0}^{1}|\dot{\bar x}_m(\bar s)|^2 d\bar s -
\int_{0}^{1}\left(\frac{\lambda}{2} |\bar x_m(\bar s)|^2 + \;
\hbox{(lower degree terms)}\right)d\bar s.
\]
If the endpoints of the curves were two fixed points, then Lemma
3.4 and Remark 3.3 in \cite{CFlS} would yield the first assertion.
Otherwise,  the result follows by connecting all the endpoints to
a fixed point by means of curves with bounded energy, and applying
previous case.

For the last assertion, just apply the first one to the sequence
of curves: $$ \hat x_m(s)=\left\{ \begin{array}{ll}
x_m(2s) & \quad\hbox{if}\;\; 0<s<\Delta_m/2 \\
x_m(2\Delta_m -2s) & \quad\hbox{if}\;\; \Delta_m/2<s<\Delta_m .
\end{array}\right.
$$
\cvd

Notice that the value of $\mu$ in previous result becomes
irrelevant (as $\epsilon>0$), but the inequality for the leading
coefficient $\lambda <\pi^2$ or $\lambda <\pi^2/4$ (the optimal
ones coming from Wirtinger's Inequality) must hold. Nevertheless,
such a bound for $\lambda$ can be avoided in the following cases.
In particular, the results are stated with $\mu=k=0$ without loss
of generality.

\begin{corollary}\label{wirtinger's2} Let $M$ be a Riemannian manifold and
$x:[0,\Delta_\infty)\rightarrow M$ a piecewise smooth curve with
 $\Delta_\infty<\infty$ and infinite energy.
Then, for $A_\Delta$ as in (\ref{des-def}):
$$ \lim_{\Delta \nearrow  \Delta_\infty} A_\Delta =
\int_{0}^{\Delta_\infty} |\dot{x}(s)|^2
ds -R\int_{0}^{\Delta_{\infty}}
|x(s)|^2 ds
= \infty. \qquad 
$$
Even more, for any $K>0$ 

 $$
 \lim_{\Delta \nearrow
\Delta_\infty} \left( A_\Delta - K |x_\Delta|^2\right) =\infty.
$$
\end{corollary}
{\it Proof.} For  $\delta \in (0,\Delta_\infty)$, put
$$x_\delta (\bar s) = x\left(\delta + (\Delta-\delta)\bar s \right) \quad \forall \bar s \in
[0,1]
$$
and
$$
\begin{array}{rl}
A_\Delta  & = A_\delta + \int_{\delta}^{\Delta} |\dot{x}(s)|^2 ds
-R\int_{\delta}^{\Delta} |x(s)|^2
ds \\
& =  A_\delta +  \frac{1}{\Delta-\delta} \int_{0}^{1}
|\dot{x}_\delta(\bar s)|^2 d\bar s -(\Delta-\delta)R \int_{0}^{1}
|x_\delta(\bar s)|^2
d\bar s
\\
& \geq  A_\delta +  \frac{1}{\Delta_\infty-\delta}
\int_{0}^{1}|\dot{x}_\delta(\bar s)|^2 d\bar s -(\Delta_\infty -\delta)R \int_{0}^{1}
|x_\delta(\bar s)|^2
d\bar s.

\end{array}
$$
Thus, the first assertion follows by taking $\delta$ close enough
to $\Delta_{\infty}$ in order to apply Theorem \ref{wirtinger's}
(with $\Delta_{m}\equiv 1$), i.e., $ \Delta_\infty -\delta < $
Min$\{1,\pi^2/4\,R\}$.

For the last part, exploiting that $R, K>0$ are arbitrary, it is
enough to check that $ \int_0^\Delta |\dot x(s)|^2 ds -
K|x_\Delta|^2 $ is lower bounded for any $K>0$. Notice that, for
$0<\Delta_0 < \Delta$:
$$
\left(|x_\Delta|-|x_{\Delta_0}|\right)^2 \leq \left(
\int_{\Delta_0}^\Delta |\dot x(s)| ds \right)^2 \leq
\left(\Delta-\Delta_0\right) \int^\Delta_{\Delta_0}|\dot x(s)|^2ds
\leq \left(\Delta_\infty-\Delta_0\right)
\int^\Delta_{\Delta_0}|\dot x(s)|^2ds.
$$
Thus, the result follows easily by taking $\Delta_0$ so that
$\Delta_\infty-\Delta_0<1/2K$.
 \cvd

\begin{proposition} \label{pwirt}
Let $M$ be a Riemannian manifold and $R_1\geq 0, R_2\in \R, 0
<\epsilon<2$. There exists $\delta
>0$, which can be taken $\delta=\infty$ if $R_1=0$,
such that
$$
\int_{0}^{\overline{\Delta}}|\dot{y}(s)|^2
ds -\int_{0}^{\overline{\Delta}}
\left( R_1 |y(s)|^2 + R_2 |y(s)|^{2-\epsilon} \right) ds > 0
$$
for all $\overline{\Delta} \in (0,\delta)$,  $y\in {\cal C}(x_0,
\bar x_0; \overline{\Delta})$ and $ x_0, \bar x_0\in M$.

\end{proposition}

{\em Proof.} For simplicity, the proof will be carried out with
$R_2=0$, being obvious the extension to the case $R_2\neq 0$.
First, putting $\tilde y (\bar s) = y(\overline{\Delta} \bar s)$:
$$
\overline{\Delta}\int_{0}^{\overline{\Delta}}|\dot y(s)|^2 ds
-\frac{\pi^2}{2\overline{\Delta}}\int_{0}^{\overline{\Delta}}
|y(s)|^2ds  
 = \int_{0}^{1}
 |\dot{\tilde y}(\bar s)|^2 d\bar s -\frac{\pi^2}{2}\int_{0}^{1}
 |\tilde y(\bar s)|^2 d\bar s  \geq 0, 
 $$ the latter by
Wirtinger's inequality. Thus,
$$
R_1\int_{0}^{\overline{\Delta}} |y(s)|^2ds \leq \frac{2R_1
\overline{\Delta}^2}{\pi^2}\int_{0}^{\overline{\Delta}}|\dot
y(s)|^2 ds,
$$
and the required inequality follows obviously if $\delta \leq \pi
/ \sqrt{2R_1}$. \cvd


 }
\end{document}